\documentclass[english,onecolumn]{article}

\usepackage[utf8]{inputenc}

\usepackage[big]{dgruyter}
\usepackage{aux}

\begin{document}
\articletype{Research Article{\hfill}Open Access}

\author*[1]{Johan de Aguas}
\affil[1]{Dpt. Mathematics, University of Oslo; Dpt. Child Health \& Development, Norwegian Institute of Public Health, Oslo, Norway; E-mail: \texttt{johanmd@math.uio.no}.}

\author[2]{Johan Pensar}
\affil[2]{Dpt. Mathematics, University of Oslo; Norway.}

\author[3]{Tomás Varnet Pérez}
\affil[3]{Dpt. Child Health \& Development, Norwegian Institute of Public Health, Oslo, Norway.}

\author[4]{Guido Biele}
\affil[4]{Dpt. Child Health \& Development, Norwegian Institute of Public Health, Oslo, Norway.}

\runningauthor{Johan de Aguas \textit{et al.}}

\title{\huge Recovery and inference of causal effects with sequential adjustment for confounding and attrition}

\runningtitle{Sequential adjustment for confounding and attrition}


\begin{abstract}{
Confounding bias and selection bias bring two significant challenges to the validity of conclusions drawn from applied causal inference. The latter can stem from informative missingness, such as in cases of attrition. We introduce the \textit{sequential adjustment criteria}, which extend available graphical conditions for recovering causal effects from confounding and attrition using sequential regressions, allowing for the inclusion of post-exposure and forbidden variables in the adjustment sets. We propose an estimator for the recovered \textit{average treatment effect}  based on \textit{targeted minimum-loss estimation}, which exhibits multiple robustness under certain technical conditions. This approach ensures consistency even in scenarios where the \textit{double inverse probability weighting} and the naïve plug-in sequential regressions approaches fall short. Through a simulation study, we assess the performance of the proposed estimator against alternative methods across different graph setups and model specification scenarios. As a motivating application, we examine the effect of pharmacological treatment for \textit{attention-deficit/hyperactivity disorder} upon the scores obtained by diagnosed Norwegian schoolchildren in national tests using observational data ($n=9\,352$). Our findings align with the accumulated clinical evidence, affirming a positive but small impact of medication on academic achievement.
}
\end{abstract}
  \journalname{Journal of Causal Inference}
  \startpage{1}
  \received{05-02-2024}
  \revised{30-09-2024}
  \accepted{02-02-2025}

  \journalyear{2025}
 
\maketitle 


\section{Introduction}


Two of the most prominent challenges to the validity of conclusions drawn from applied causal inference are \textit{confounding bias} and \textit{selection bias}. The former arises from unaccounted flows of statistical information between the exposure and the outcome under study, thereby making it difficult to disentangle the target causal effect from other correlational associations in the system \citep{PearlRobins,greenland2003,vanderweele2013d}. The latter is an umbrella term that encompasses various types of biases that emerge when inference for a causal or statistical query for a target population is performed using a non-representative subpopulation or a sample with preferential characteristics \citep{hernan2004structural,evolutionSelection,colliderStratification}. 

Selection bias can be introduced through \textit{informative missingness}, which refers to situations wherein pieces of relevant information regarding some units from the target population are systematically missing, censored or coarsened  \citep{westreich2012}. Conducting an analysis based solely on complete cases may lead to erroneous inference if the causes for missingness also influence or are influenced by the outcome \citep{Hernan2017}, or when the exposure‐outcome association shifts for the selected relative to the missing units \citep{Australia}. Furthermore, results may lack a causal interpretation even within the selected subpopulation, due to paths of spurious associations or non-causal contrasts induced by conditioning on selection \citep{Bareinboim12,mathur2023}. 

A causal effect is said to be \textit{recoverable} if it is uniquely computable as a functional of a given positive observed data distribution  \citep{Bareinboim_Tian_Pearl_2014}. Unlike the problem of identification from confounding bias, recovery from selection bias typically cannot be resolved through a random treatment assignment mechanism, so it requires a graphical model approach for resolution \citep{hernan2004structural,Bareinboim12}. One strategy consists of incorporating the \textit{selection} or \textit{missingness mechanisms} into the systems's causal graph to explicitly represent the mix of exogenous and endogenous factors that determine participation or censoring \citep{geneletti2009adjusting,didelez2010graphical,evolutionSelection}. These augmented graphs are known as selection diagrams or as \textit{$m$-graphs} in the context of missing data. Graphical criteria applied to these $m$-graphs can enable the recovery of statistical and causal parameters even in complicated settings that extend beyond the conventional assumption of a \textit{missing at random} (MAR) model \citep{Mohan2013,mgraphs,NabiSemi}. 

In the context of the problem of missing outcome data, often referred to as \textit{attrition} or \textit{loss-to-follow-up} \citep{attrition,biele2019bias}, significant attention has been given to establishing sufficient recovery criteria for causal effects in scenarios where the missingness mechanism is influenced by the exposure, pre-exposure variables, and non-\textit{forbidden} variables, i.e., those not descending from intermediate nodes in causal pathways from the exposure to the outcome \citep{ShpitserRobin10,Oset}. Solutions for additive, cumulative, and quantile causal effects have been advanced, including those based on sequential factorizations and covariate adjustment \citep{Mohan2014} ---such as the \textit{generalized adjustment criteria} (GAC) \citep{Correa_Tian_Bareinboim_2018} and the \textit{$m$-adjustment} criteria \citep{saadati2019adjustment}---, through \textit{double inverse probability weighting} (DIPW) \citep{HuberDW}, and via techniques that incorporate both covariate adjustment and sample weighting \citep{DREE,Wei2022,dw,Tang2024}. The latter type of approach has become favored for its \textit{double robustness} property, which guarantees consistency of the estimator even when either the outcome model or the propensity score models are misspecified, but not both simultaneously \citep{DRbookVariance,mcisaac2017,AIPW2017}.

In contrast, settings that involve forbidden nodes or possible descendants of the exposure have garnered less attention, despite their potential to more accurately model attrition mechanisms in various empirical contexts. In these cases, recoverability can be assessed via the identification algorithm by \citet{Bhattacharya2020} or the search-based heuristic developed by \citet{tikka2021causal}. Conditions outlined by \citet{Mohan2014} for the \textit{simple attrition} case effectively cover this setting under an exogenous or randomized treatment assignment. However, to the best of our knowledge, only \citet{Huber} provides itemized sufficient criteria for recovering additive causal effects in cases of attrition with \textit{selection-outcome intermediate confounding} (SOIC), where a descendant of the exposure confounds the outcome and its selection indicator.

This article makes two primary contributions. First, we provide sufficient conditions that enable the recovery of causal effects from attrition in structures where existing criteria fall short. Second, we introduce the \textit{targeted sequential regression} (TSR) estimator for the \textit{average treatment effect} (ATE) when the recovered estimand involves two sequential regressions models.

For this, we introduce a graph-based \textit{$s$-admissible pair} of covariate sets, which satisfy the \textit{sequential adjustment criteria} (SAC). These criteria are itemized conditions for recovering the interventional distribution and additive causal effects from confounding and attrition biases in tandem. Our approach is grounded in the \textit{structural causal model} (SCM) framework pioneered by Pearl \citep{PearlCausality}, which allows us to directly translate recovery criteria into conditions on mutilated graphs \citep{mgraphs}. When fulfilled, the SAC provides solutions to settings involving SOIC and other attrition structures. While similar to conditions by \citet{Huber} in addressing the SOIC problem, our method allows for recovery in a broader scope of semi-Markovian structures, where their approach is limited by chronological constraints in the adjustment rules. In addition, our approach can aid in identifying and mitigating potential pitfalls such as \textit{overadjustment} \citep{over1,over2,over3} and \textit{collider bias} \citep{mbias2}. 



The TSR estimator is developed to estimate the ATE in situations where the SAC yields an estimand involving two sequential regression models. It is built on the \textit{targeted minimum-loss estimation} (TMLE) framework \citep{vanderLaanRubin2006,gruber2009targeted}, and is \textit{multiplyrobust} under certain conditions. Specifically, it remains consistent if at least one of these conditions holds: \textit{(i)} the sequential regression models are correctly specified, \textit{(ii)} the exposure and selection propensity score models are correctly specified, and \textit{(iii)} the exposure propensity score model and the mean-imputation model are correctly specified. In a sense, the TSR estimator capitalizes on the collective robustness conditions of regression-based, IPW-based, and imputation-based solutions. Our approach thus broadens the scope of scenarios where a TMLE procedure can yield a consistent estimate of a causal effect in the presence of missing data \citep{Dashti2021,phd}. It offers a superior bias-variance tradeoff, and its uncertainty quantification demonstrates greater robustness to model misspecification than other approaches. As a result, it produces confidence intervals with more robust coverage probability, leading to more reliable statistical inference.

We apply the developed procedures to the problem of estimating the causal effects of pharmacological treatment for \textit{attention-deficit/hyperactivity disorder} (ADHD) upon the scores obtained by diagnosed Norwegian schoolchildren in national tests. Exemptions and abstentions from these tests may be influenced by endogenous factors, including potential adverse effects of stimulant medication such as insomnia and weight loss \citep{graham2008adverse}. Our devised theoretical and methodological tools are well suited for this applied research question, as adverse drug reactions are commonly indicative of different forms of attrition \citep{sant2023impact, bikoro2020incidence}.

In \cref{secPrelim}, we introduce essential notation and tools from the SCM framework, and several graphical criteria for recovering causal effects under attrition. In \cref{secProblem,secSAC,secPlugin}, we discuss the limitations of these methods and present the SAC alongside several plug-in estimators. Under these criteria, \cref{secidTMLE} outlines a TMLE-based procedure for estimating the recovered ATE via a sequential regression strategy. We apply this procedure to simulated data in \cref{secSimulation}, comparing it with other common estimators across simple graphical scenarios. In \cref{secApplication}, we use real-world observational data to estimate the causal effects of pharmacological treatment for ADHD upon the scores obtained by diagnosed Norwegian schoolchildren in national tests. Finally, we present summarizing conclusions in \cref{secDiscussion}.


\section{Preliminaries}
\label{secPrelim}

We present in this section the theory, notation, and definitions pertinent to the recovery of causal effects, which are later employed in the formulation of the SAC in \cref{secSAC}.

\subsection{Genealogic sets and graph surgery}
Given a \textit{directed acyclic graph} (DAG) $\mathcal{G}$  on a set of nodes $\mathcal{V}$, and a subset of them $A\subseteq\mathcal{V}$, let  $\an(A;\mathcal{G})$ and $\de(A;\mathcal{G})$ denote, respectively, the ancestors and descendants of $A$ in $\mathcal{G}$. Their \textit{inclusive} ancestors and descendants are denoted with $\An(A;\mathcal{G})$ and $\De(A;\mathcal{G})$, respectively, and include nodes $A$ themselves \citep{gID2020}. The non-descendants of $A$ in $\mathcal{G}$ are  $\nd(A;\mathcal{G}):=\mathcal{V}\,\setminus\De(A;\mathcal{G})$. 

For two disjoint subsets of nodes $A,B\subset\mathcal{V}$, the \textit{edge-mutilated} graph $\mathcal{G}\,\setminus (B;A)$ is the graph resulting from deleting in $\mathcal{G}$ all arrows coming out of nodes in $B$ that go into nodes in $A$. Denote in bar notation:
\begin{equation}
    \mathcal{G}[\overline{A}] := \mathcal{G}\,\setminus (\mathcal{V};A) \qquad\text{ and }\qquad \mathcal{G}[\underline{B}]:= \mathcal{G}\,\setminus (B;\mathcal{V}).
\end{equation}

Let $\cn(A,Y;\mathcal{G})$ be the set of \textit{proper causal nodes} with respect to disjoint $A,Y\subset\mathcal{V}$ in $\mathcal{G}$, i.e., the set of nodes in non-self-intersecting directed paths from $A$ to $Y$, excluding $A$ \citep{ShpitserRobin10,Oset2}: 
\begin{equation}
    \cn(A,Y;\mathcal{G}) := \de(A;\mathcal{G}[\overline{A}])\cap 
     \An(Y;\mathcal{G}[\underline{A}]).
\end{equation}
We use $\fb(A,Y;\mathcal{G})$ to denote the set of \textit{forbidden nodes} with respect to disjoint $A,Y\subset\mathcal{V}$ in $\mathcal{G}$, which is the set of proper causal nodes and their descendants, along with the exposure \citep{Oset}: 
\begin{equation}
    \fb(A,Y;\mathcal{G}) := \De\left(\cn(A,Y;\mathcal{G});\mathcal{G}\right)\cup A.
\end{equation}

This is also referred as the DPCP set: the \textit{descendants of nodes in proper causal paths}, when it does not include $A$ \citep{vanderzander2014}. We represent with $\mathcal{G}[A\mid Y]$ the \textit{proper  backdoor graph} with respect to disjoint $A,Y\subset\mathcal{V}$. This is the graph obtained by removing from $\mathcal{G}$ all arrows coming from nodes in $A$ and entering any proper causal node \citep{vanderzander2014}:
\begin{equation}
    \mathcal{G}[A\mid Y] := \mathcal{G}\,\setminus (A;\cn(A,Y;\mathcal{G})).
\end{equation}

Finally, we use $A\indep_d B$ to denote that $A$ and $B$ are $d$-separated in a graph $\mathcal{G}$ \citep{pearl1988probabilistic}.

\subsection{Structural causal models}

An SCM is a tuple $\mathcal{M}=(\mathcal{V},\mathcal{U},\mathcal{G},\mathcal{F},P_\mathcal{U})$. Here, $\mathcal{V}$ and $\mathcal{U}$ are finite sets of relevant variables and of latent background noises, respectively; $\mathcal{G}$ is a DAG on $\mathcal{V}$; and  $P_\mathcal{U}$ is a probability measure for $\mathcal{U}$. Finally, $\mathcal{F}=\{f_V\}_{V\in\mathcal{V}}$ is an indexed collection of measurable functions specifying the direct causal mechanisms, i.e., for every $V\in\mathcal{V}$ there is a $U_V\in\mathcal{U}$ and a function $f_V:\supp \pa(V;\mathcal{G})\times \supp U_V\rightarrow\supp V$, such that $V=f_V(\pa(V;\mathcal{G}),U_V)$ almost surely, where $\pa(V;\mathcal{G})$ represents the parent nodes of $V$ in $\mathcal{G}$ \citep{PearlCausality}. 

Let disjoint variables $A,Y\subset\mathcal{V}$ denote the \textit{exposure} and the \textit{outcome}, respectively. The \textit{unit-level counterfactual} or \textit{potential outcome} $Y^{a}(u)$ is the value $Y$ takes after an intervention that sets $A$ to the fixed value $a\in\supp A$ for the individual context $\mathcal{U}=u$ in the SCM $\mathcal{M}$, and propagates this change to the other variables by applying the mechanisms $\mathcal{F}$ following a topological order of $\mathcal{G}$. Potential outcomes respect the consistency axiom, i.e., $A(u)=a,Y(u)=y\Rightarrow Y^{a}(u)=y$   \citep{robinsConsistency,PearlCausality}. The induced population-level distribution, named the \textit{interventional distribution}, is given by:
\begin{equation}
    p(y\mid\doo(A=a)) := p_{Y^{a}}(y)=\int\I\left\{u\in \mathcal{U}^a[y]\right\}\,\dd P(u),
\end{equation}
where $\mathcal{U}^a[y]=\{u\in\supp\, \mathcal{U} : Y^{a}(u)=y \}$ is the inverse image of $y\in\supp Y$ under $Y^{a}(\cdot)$ \citep{BareinboimHierarchy2022}.  

The ATE, $\psi$, is the most commonly investigated \textit{causal effect} or \textit{query}. For a binary exposure $A$, it corresponds to a difference functional of the interventional distribution:
\begin{equation}
    \psi := \Delta_a \E\left[Y\mid\doo(A=a) \right],
\end{equation}

\noindent where $\Delta_a$ denotes the difference operator relative to the binary argument $a$. This is, for any measurable function $f$ of $x\in\mathcal{X}$ and $a\in\{0,1\}$, one has $\Delta_af(x,a):=f(x,1)-f(x,0)$.

\subsection{Causal effect recovery from missing data and attrition}

Let $\mathcal{M}=(\mathcal{V},\mathcal{U},\mathcal{G},\mathcal{F},P_\mathcal{U})$ be an SCM, and  $\mathcal{V}_o\cup\mathcal{V}_m\cup\mathcal{R}$ be a partition of variables $\mathcal{V}$, where $\mathcal{V}_o$ contains fully observed variables, $\mathcal{V}_m$ contains variables affected by missingness, and $\mathcal{R}=\{R_V : V\in\mathcal{V}_m\}$ collects the missingness indicators. That is, if $V$ is observed for an individual unit, then $R_V=1$ for the same unit; and if it is missing, then $R_V=0$. In this augmented system, the causal graph $\mathcal{G}$ is called a missingness graph, or \textit{$m$-graph}, as it involves the selection indicators $\mathcal{R}$ in addition to the \textit{substantive variables} $\mathcal{V}_o\cup\mathcal{V}_m$. It is typically assumed that $\mathcal{R}$ has no substantive descendants in $\mathcal{G}$ \citep{mgraphs,nabi2022causal}. Let us denote with $V^\dagger$ the \textit{proxy} for $V\in\mathcal{V}_m$, such that $V^\dagger=V$ almost surely when $R_V=1$, and it takes the empty value $V^\dagger=\oldempty$ otherwise. The distribution $P_\mathcal{V}$ is known as the \textit{full data distribution}, while $P_{\mathcal{V}^\dagger}$, with $\mathcal{V}^\dagger=\mathcal{V}_o\cup\mathcal{V}_m^\dagger\cup\mathcal{R}$, is the \textit{observed data distribution}.

A causal query $\Psi$ is  \textit{(nonparametric) recoverable} from observational data if it outputs the same value $\psi$ for all models in the class of SCMs $\mathfrak{M}$ sharing the same $m$-graph $\mathcal{G}$  and the same positive observed data distribution $P$, i.e., $\forall\mathcal{M}\in\mathfrak{M}(\mathcal{G},P),\, \Psi[\mathcal{M}]=\psi$. That is, the query is \textit{uniquely computable} as a functional of the graph and the data \citep{Bareinboim_Tian_Pearl_2014}, yielding a statistical \textit{estimand}. Various sets of recoverability conditions exist, with potentially different feasibility and output estimands  \citep{Correa_Tian_Bareinboim_2019}, and some may apply only to specific parameterizations of the target query \citep{Pearl2013simple}. Specifically, the recoverability of interventional distributions can be evaluated using the identification algorithm proposed by \citet{Bhattacharya2020} or the search-based heuristic developed by \citet{tikka2021causal}, though neither approach has been proven to be complete.

Consider the $m$-graph depicted in  \cref{fig1a}, which encodes a MAR model. Here, $A$ denotes a binary point exposure, $Y$ represents the outcome, and $R_Y$ is the binary selection indicator for the outcome, where $R_Y=1$ indicates that the outcome is observed. Recovery of the ATE is achievable under certain positivity conditions using a ratio factorization of the query \citep{Mohan2014}, leading to the DIPW estimand:
\begin{equation}\label{eqIPW1}
    \psi = \Delta_a\E \left[\frac{\mathbb{I}(A=a)\cdot R_Y\cdot Y}{\pr(A=a\mid B_1)\cdot\pr(R_Y=1\mid B_1,C_1,A)} \right]. 
\end{equation}

Recovery is also feasible via generalized adjustment \citep{Correa_Tian_Bareinboim_2018}, or \textit{$g$-adjustment}, by leveraging a valid sequential factorization of the query \citep{Mohan2014} and an \textit{admissible} adjustment set, $\{B_1,C_1\}$ in this scenario, leading to a regression-based estimand:
\begin{equation}
    \psi =\Delta_a\E\left[\E\left[Y\mid B_1,C_1,A=a,R_Y=1\right]\right].
\end{equation}




\section{Setting and problem formulation}
\label{secProblem}

The setting considered in this work corresponds to a population affected by attrition, and can be formalized as follows. 
    Let $A\in\{0,1\}$ denote a binary point exposure, $Y\in\R$ a continuous outcome, $\mathcal{V}=\mathcal{V}_0\cup \mathcal{V}_1$ a set containing chronological pre-exposure covariates $\mathcal{V}_0$ realized in time before the exposure, and chronological post-exposure covariates $\mathcal{V}_1$ realized in time after the exposure and before the outcome. Let $R_Y\in\{0,1\}$ be the selection indicator of the outcome, and $\mathcal{G}$ be an $m$-graph on $\mathcal{V}'=\mathcal{V}\cup\{A,Y,R_Y\}$, where $R_Y$ is a childless node. Finally, let $P_{\rm{obs}}$ denote the observed data distribution, i.e., the joint distribution of the vector $(\mathcal{V},A,Y^\dagger,R_Y)$ in the reference population.


\begin{figure}[t]
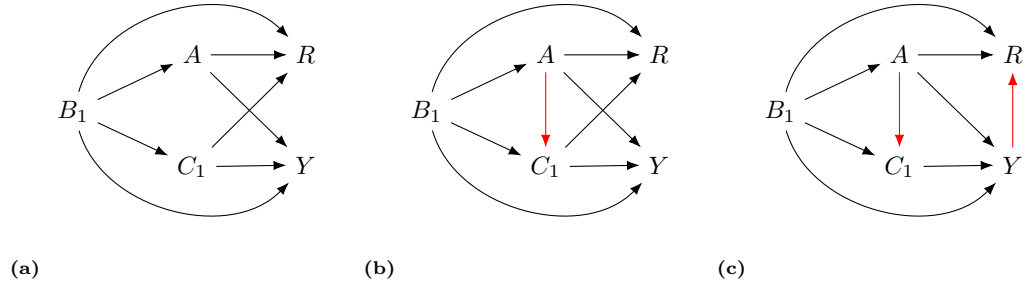

\centering
\begin{subfigure}{.3\textwidth}
  \centering
\tikz[scale=0.8, transform shape]{
    \node (h) {$B_1$};
    \node (a) [right = of h, yshift=0.75cm] {$A$};
    \node (s) [right = of a] {$R_Y$};
    \node (m) [below = of a] {$C_1$};
    \node (y) [below = of s] {$Y$};

    \path (a) edge[style=directed] (y);
    \path (h) edge[style=directed] (a);
    \path (h) edge[style=directed] (m);
    \path (m) edge[style=directed] (y);
    \path (m) edge[style=directed] (s);
    \path (a) edge[style=directed] (s);
    \path (h) edge[style=directed,bend right=60] (y);
    \path (h) edge[style=directed,bend left=60] (s);
}
    \caption{}
    \label{fig1a}
\end{subfigure}%
\begin{subfigure}{.3\textwidth}
  \centering
\tikz[scale=0.8, transform shape]{
    \node (h) {$B_1$};
    \node (a) [right = of h, yshift=0.75cm] {$A$};
    \node (s) [right = of a] {$R_Y$};
    \node (m) [below = of a] {$C_1$};
    \node (y) [below = of s] {$Y$};

    \path (a) edge[style=directed] (y);
    \path (h) edge[style=directed] (a);
    \path (a) edge[style=directed, red] (m);
    \path (h) edge[style=directed] (m);
    \path (m) edge[style=directed] (y);
    \path (m) edge[style=directed] (s);
    \path (a) edge[style=directed] (s);
    \path (h) edge[style=directed,bend right=60] (y);
    \path (h) edge[style=directed,bend left=60] (s);
}
    \caption{}
    \label{fig1b}
\end{subfigure}%
\begin{subfigure}{.3\textwidth}
  \centering
\tikz[scale=0.8, transform shape]{
    \node (h) {$B_1$};
    \node (a) [right = of h, yshift=0.75cm] {$A$};
    \node (s) [right = of a] {$R_Y$};
    \node (m) [below = of a] {$C_1$};
    \node (y) [below = of s] {$Y$};

    \path (a) edge[style=directed] (y);
    \path (h) edge[style=directed] (a);
    \path (a) edge[style=directed, red] (m);
    \path (h) edge[style=directed] (m);
    \path (m) edge[style=directed] (y);
    \path (a) edge[style=directed] (s);
    \path (h) edge[style=directed,bend right=60] (y);
    \path (h) edge[style=directed,bend left=60] (s);
    \path (y) edge[style=directed, red] (s);
}
    \caption{}
    \label{fig1c}
\end{subfigure}
\caption{$m$-graphs with different recoverability results for the interventional distribution: (a) recoverable via DIPW and via $g$-adjustment (type 2), (b) recoverable via DIPW and via the SAC, and (c) not recoverable by either due to self-selection (MNAR).}
\label{fig1}
\end{figure}

The $m$-graph depicted in \cref{fig1b} illustrates a MAR model involving SOIC, a form of time-dependent or dynamic confounding \citep{Rotnitzky,Huber}. Here, the selection indicator $R_Y$ and the outcome $Y$ are confounded by a variable $C_1$ that is influenced by the exposure $A$. Recovery is still achievable through DIPW with the same recovered estimand as in \cref{eqIPW1}. In contrast, recovery via $g$-adjustment is no longer possible, as there is no covariate adjustment set devoid of $C_1$ that fulfills all the required conditions. This issue is known as the \textit{type-$m$ insufficiency problem} \citep{mathur2023}. One consequence is the lack of recoverability via a single regression using pre-exposure adjustment sets. In addition, the selected subpopulation does not constitute a causal stratum from the reference population, and the query $\xi:=\Delta_a\E\left[Y\mid \doo(A=a),R_Y=1\right]$ fails to define a valid causal contrast \citep{Bareinboim12}. The alternative valid contrast  $\zeta:=\Delta_a\E\left[Y^a\mid R^0_Y=1\right]$ is a pure counterfactual query that escapes experimental means of identification, as it conditions on the potential outcome $R^0_Y$ after intervention $\doo(A=0)$ \citep{robins2007discussions,dawid2012imagine}.

This explains why, in practice, the DIPW approach is favored for estimating causal effects in cases of missing, censored, or coarsened outcome data in observational studies, as it can allow the inclusion of post-exposure variables to model the selection probability. Nevertheless, IPW-based estimators often face challenges related to instability, sensitivity to near positivity violations, and the generation of wider confidence intervals \citep{instability,seaman2013review,stableIPW}. They also complicate valid uncertainty quantification, particularly within the Bayesian framework and other likelihood-centered approaches \citep{wasserman}.

Recently, \citet{Huber} introduced itemized sufficient conditions for recovering expected potential outcomes under attrition with SOIC, within the framework of Rubin's causal model \citep{rubin74}. However, these criteria rely on the chronological order of variables and are not constructive, as they do not specify a method to build or identify the necessary adjustment sets from a given causal graph. Constructive approaches facilitate the identification of one or multiple admissible adjustment sets from a given causal graph, if any exist \citep{vanderzander2014,Correa_Tian_Bareinboim_2018,Oset}, can aid in evaluating issues such as \textit{overadjustment} \citep{over1,over2,over3} and potential \textit{collider bias} \citep{mbias2}, and in elucidating \textit{optimal} adjustment strategies \citep{Oset2}. In addition, itemized criteria enhance transparency by making the assumptions about the data-generating process more explicit, allowing practitioners to independently justify each component \citep{mgraphs}. To fill this gap, we present the SAC, which enlist sufficient graphical conditions for recovering causal effects under attrition.


\section{Sequential adjustment criteria}
\label{secSAC}


In this section, we define an \textit{$s$-admissible pair} as the combination of a valid inner covariate set $Z$ and an outer covariate set $W$, which can be used to recover the interventional distribution and additive causal effects through sequential regressions. We also provide an analysis of minimal $s$-admissible pairs and their modeling implications in the context of some selected $m$-graphs.

\begin{definition}[$s$-admissibility]\label{defSAC}
Let $A,Y,R_Y,\mathcal{V}$, and $\mathcal{G}$ be described as in \cref{secProblem}. Let $W,Z$ be disjoint sets of variables in $\mathcal{V}$. We say $(W;Z)$ is an $s$-admissible pair relative to $(A;Y)$ in $\mathcal{G}$ if it fulfills the SAC, given by:
\begin{enumerate}[leftmargin=30pt]
    \item $W$ contains no forbidden nodes: $W\cap\fb(A,Y;\mathcal{G})=\emptyset$,
    \item $W$ $d$-separates $Y$ from $A$ in the proper backdoor graph: $Y\indep_d A\mid W$ in $\mathcal{G}[A\mid Y]$,
    \item $(W,A,Z)$ $d$-separates $Y$ from $R_Y$ in the original $m$-graph: $Y\indep_d R_Y\mid W,A,Z$ in $\mathcal{G}$.
\end{enumerate}
\end{definition}

In Markovian SCMs, where all exogenous noise variables $\mathcal{U}$ are mutually independent, and in settings involving a point exposure and no self-selection ($R_Y\notin\de(Y;\mathcal{G})$), $s$-admissible pairs are guaranteed to exist. This is because one can set $W = \pa(Y; \mathcal{G}) \setminus \fb(A, Y; \mathcal{G})$ and $Z = (\pa(Y; \mathcal{G}) \cap \fb(A, Y; \mathcal{G}))\setminus A$, which will satisfy the SAC.


\begin{definition}[Minimal $s$-admissible pair]\label{defMin}
    Let $(W;Z)$ be an $s$-admissible pair relative to $(A;Y)$ in $\mathcal{G}$. $(W;Z)$ is minimal if removing any variable from $W$ or $Z$ renders the resulting pair no longer $s$-admissible.
\end{definition}

\begin{theorem}\label{theoSAC}
Let $(W;Z)$ be an $s$-admissible pair relative to $(A;Y)$ in $\mathcal{G}$, and $P$ be a positive distribution for the vector $O=(W,A,Z,Y^\dagger,R_Y)$. Then, the interventional distribution and the ATE are recovered as follows:
\begin{align}\label{eqSAC0}
     p(y\mid\doo(A=a)) &= \begin{cases}
      \E_W\,p(y\mid W,A=a,R_Y=1) & \text{if } Z=\emptyset,\\
     \E_W\E_{Z\mid W,A=a}\, p(y\mid W,A=a,Z,R_Y=1) & \text{if } Z\neq\emptyset,\\
\end{cases}\\
     \psi &= \begin{cases}
     \E_W\Delta_a\E\left[Y\mid W,A=a,R_Y=1\right]  & \text{if } Z=\emptyset,\\
     \E_W\Delta_a\E_{Z\mid W,A=a}\, \E\left[Y\mid W,A=a,Z,R_Y=1\right]  & \text{if } Z\neq\emptyset.
     \end{cases}
     \label{eqSAC}
\end{align}
\end{theorem}

A proof is provided in \cref{appendixA1}. 

To avoid a cluttered notation, sequential expectations are presented in a streamlined notation without nested brackets. Therefore, expressions such as $\E_W\Delta_a\E_{Z\mid W,A=a}\, \E\left[Y\mid W,A=a,Z,R_Y=1\right]$ must be interpreted as $\E\left(\Delta_a\E\left\{\E\left[Y\mid W,A=a,Z,R_Y=1\right] \mid W,A=a\right\} \right)$.

When $Z=\emptyset$, \cref{theoSAC} produces the estimand given by $g$-adjustment (type-2) \citep{Correa_Tian_Bareinboim_2018}. If $W=\emptyset$ in this context as well, the setting corresponds to an exogenous treatment assignment, and the recovered ATE reduces to $\psi=\Delta_a\E\left[Y\mid A=a,R_Y=1\right]$. When $Z\neq\emptyset$, two sequential adjustment sets may be required. If $W=\emptyset$ in this context, the setting corresponds to the \textit{simple attrition} case \citep{Mohan2014}, with recovered ATE  $\psi=\Delta_a\E_{Z\mid A=a}\, \E\left[Y\mid A=a,Z,R_Y=1\right]$.

\begin{figure}[t]
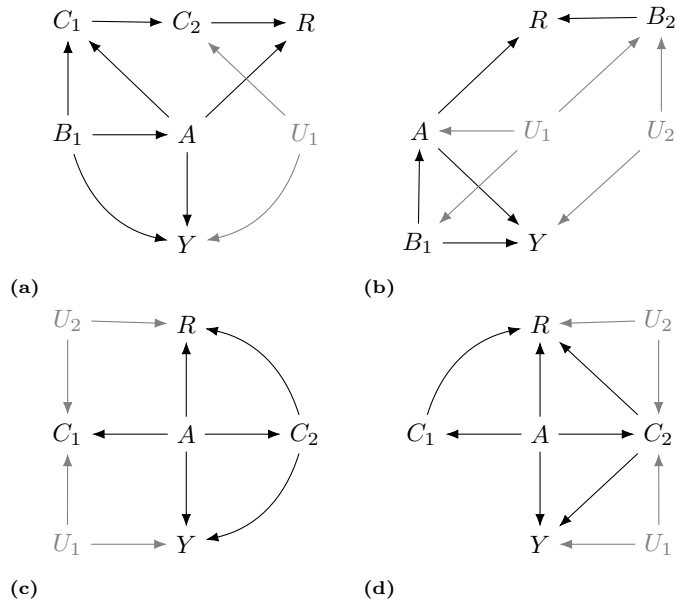

\centering
\begin{subfigure}{.225\textwidth}
  \centering
\tikz[scale=0.8, transform shape]{
    \node (w) {$B_1$};
    \node (a) [right = of w] {$A$};
    \node (z) [right = of a, gray] {$U_1$};
    \node (r) [above = of a] {$C_2$};
    \node (y) [below = of a] {$Y$};
    \node (z1) [above = of w] {$C_1$};
    \node (u4) [above = of z] {$R_Y$};

    \path (w) edge[style=directed] (a);
    \path (a) edge[style=directed] (u4);
    \path (w) edge[style=directed] (z1);
    \path (a) edge[style=directed] (y);
    \path (w) edge[style=directed,bend right=30] (y);
    \path (a) edge[style=directed] (z1);
    \path (z1) edge[style=directed] (r);
    \path (z) edge[style=directed, gray] (r);
    \path (z) edge[style=directed, gray, bend left=30] (y);
    \path (r) edge[style=directed] (u4);
}
    \caption{}
    \label{fig2a}
\end{subfigure}%
\begin{subfigure}{.225\textwidth}
  \centering
\tikz[scale=0.8, transform shape]{
    \node (w) {$A$};
    \node (a) [right = of w, gray] {$U_1$};
    \node (z) [gray, right = of a] {$U_2$};
    \node (r) [above = of a] {$R_Y$};
    \node (y) [below = of a] {$Y$};
    \node (u4) [above = of z] {$B_2$};
    \node (w1) [left = of y] {$B_1$};

    \path (a) edge[style=directed, gray] (w);
    \path (w1) edge[style=directed] (w);
    \path (w1) edge[style=directed] (y);
    \path (a) edge[style=directed, gray] (u4);
    \path (w) edge[style=directed] (r);
    \path (w) edge[style=directed] (y);
    \path (a) edge[style=directed, gray] (w1);
    \path (z) edge[style=directed, gray] (y);
    \path (z) edge[style=directed, gray] (u4);
    \path (u4) edge[style=directed] (r);
}
    \caption{}
    \label{fig2b}
\end{subfigure}
\begin{subfigure}{.225\textwidth}
  \centering
\tikz[scale=0.8, transform shape]{
    \node (w) {$C_1$};
    \node (a) [right = of w] {$A$};
    \node (z) [right = of a] {$C_2$};
    \node (r) [above = of a] {$R_Y$};
    \node (y) [below = of a] {$Y$};
    \node (u1) [gray, above = of w] {$U_2$};
    \node (u2) [gray, left = of y] {$U_1$};

    \path (a) edge[style=directed] (w);
    \path (a) edge[style=directed] (z);
    \path (a) edge[style=directed] (r);
    \path (a) edge[style=directed] (y);
    \path (z) edge[style=directed,bend right=30] (r);
    \path (z) edge[style=directed,bend left=30] (y);
    \path (u1) edge[style=directed, gray] (w);
    \path (u1) edge[style=directed, gray] (r);
    \path (u2) edge[style=directed, gray] (w);
    \path (u2) edge[style=directed, gray] (y);
}
    \caption{}
    \label{fig2c}
\end{subfigure}
\begin{subfigure}{.225\textwidth}
  \centering
\tikz[scale=0.8, transform shape]{
    \node (w) {$C_1$};
    \node (a) [right = of w] {$A$};
    \node (z) [right = of a] {$C_2$};
    \node (r) [above = of a] {$R_Y$};
    \node (y) [below = of a] {$Y$};
    \node (u2) [gray, right = of y] {$U_1$};
    \node (u4) [gray, above = of z] {$U_2$};

    \path (a) edge[style=directed] (w);
    \path (a) edge[style=directed] (z);
    \path (a) edge[style=directed] (r);
    \path (a) edge[style=directed] (y);
    \path (w) edge[style=directed,bend left=30] (r);
    \path (z) edge[style=directed] (r);
    \path (z) edge[style=directed] (y);
    \path (u2) edge[style=directed, gray] (z);
    \path (u2) edge[style=directed, gray] (y);
    \path (u4) edge[style=directed, gray] (z);
    \path (u4) edge[style=directed, gray] (r);
}
    \caption{}
    \label{fig2d}
\end{subfigure}
\caption{$m$-graphs with minimal $s$-admissible pairs: (a) $(\{B_1,C_1,C_2\};\emptyset)$ and $(\{B_1\};\{C_2\})$, (b) $(\{B_1\};\{B_2\})$,  (c) $(\emptyset;\{C_2\})$, and (d) there is no $s$-admissible pair. In all these graphs, $U_1$ and $U_2$ are latent variables.}
\label{fig2}
\end{figure}

Similar to the \textit{extended} admissibility conditions for the backdoor criterion, the specific components of an $s$-admissible pair are not confined to chronological pre- and post-exposure periods \citep{PearlPT}. This differs from the criteria by \citet{Huber}, where the inner adjustment set is limited to post-exposure variables (i.e., $Z\subseteq\mathcal{V}_1$) and the outer adjustment set is restricted to variables measured before the exposure (i.e., $W\subseteq\mathcal{V}_0$). Such constraints can lead to unnecessary sequential adjustment or even to non-recoverability in some semi-Markovian SCMs. Consider, for example, the $m$-graph shown in \cref{fig1a} with $\mathcal{V}_0=\{B_1\}$ and $\mathcal{V}_1=\{C_1\}$. In this case, both $(\{B_1,C_1\};\emptyset)$ and $(\{B_1\};\{C_1\})$ qualify as minimal $s$-admissible pairs, yet only the latter adheres to the chronological constraints. The former yields a simpler recovered estimand, as per \cref{theoSAC}, requiring a single regression model when the outermost expected value is treated as fully nonparametric. This can be regarded as a special form of overadjustment \citep{over1,over2,over3}, as the latter uses more regression models than the minimum required, potentially impacting the computational efficiency and the robustness characteristics of the resulting estimators. A more appropriate criterion for comparing different $s$-admissible pairs can be established by invoking the asymptotic variance of their respective plug-in estimators, as proposed by \citet{Oset2} and \citet{Rotnitzky} for settings without attrition. We defer this analysis to future work.

In semi-Markovian SCMs, the first component of an $s$-admissible pair may include descendants of the exposure. For instance, in the $m$-graph shown in \cref{fig2a}, where $U_1$ is a latent variable, both $(\{B_1, C_1, C_2\}; \emptyset)$ and $(\{B_1\}; \{C_2\})$ constitute minimal $s$-admissible pairs. Yet again, the former produces a simpler single-regression estimand, despite violating chronological constraints. In other cases, enforcing these constrains may result in non-recoverability. In the $m$-graph depicted in \cref{fig2b}, with $\mathcal{V}_0 = \{B_1, B_2\}$ and $\mathcal{V}_1 = \emptyset$, the pair $(\{B_1\}; \{B_2\})$ is $s$-admissible, even though $B_2$ is a pre-exposure variable. If the second component is confined to post-exposure variables only, the interventional distribution of $Y$ after an intervention on $A$ becomes non-recoverable, as $B_2$ alone is not allowed in the first component.

More notably, beyond concerns about the enumeration of minimal $s$-admissible pairs and the impacts of overadjustment, improper adjustment can introduce bias. It is known that adjusting for all chronological pre-exposure covariates can lead to bias in some causal strutures involving \textit{collider paths} \citep{mbias2,cinelliControls,hunermund2023double}. Similarly, in a setting with attrition, adjusting for all potential post-exposure covariates related to the selection mechanism can produce biased estimands. This is exemplified in the $m$-graph in  \cref{fig2c}, wherein the only valid $s$-admissible pair is $(\emptyset;\{C_2\})$. Although $C_1$ is a descendant of the exposure $A$ and is associated with both the outcome $Y$ and its selection indicator $R_Y$, including it in the second component of the adjustment pair would induce bias, as it opens a collider path between $Y$ and $R_Y$. Collider structures can indeed render recovery infeasible via the SAC. In the case depicted in \cref{fig2d}, no $s$-admissible pair exists due to a \textit{butterfly structure} \citep{mbias2} that creates an \textit{inducing path} between outcome $Y$ and the indicator $R_Y$ \citep{inducing}.



\section{Naïve plug-in estimators for the ATE}
\label{secPlugin}

Let $(W;Z)$ be an $s$-admissible pair with $Z\neq\emptyset$, and $P$ denote the observed data distribution. Then, by \cref{theoSAC}, the ATE is recovered as $\psi=\Psi[P]$, with:
\begin{equation}\label{eqSformula}
    \Psi[P] = \E_W\Delta_a\E_{Z\mid W,A=a}\, \E\left[Y\mid W,A=a,Z,R_Y=1\right], 
\end{equation}

\noindent which we refer to as the \textit{$s$-formula}. The $s$-formula is an application of a recovering functional $\Psi:\mathfrak{P}\rightarrow\R$ that takes a positive observed data distribution in a semiparametric model $\mathfrak{P}$ and returns a real value.

If models for the expected outcome within the selected subpopulation and for the conditional distribution of inner covariates $Z$ are specified, then a plug-in estimator for the ATE, $\widehat{\psi}_{\text{CD}}$, can be constructed using $n$ i.i.d. samples from the observed data distribution $\{O_i\}_{i=1}^n\sim P$ as follows: 
\begin{align}\label{eqQ0}
    \widehat{Q}_1(W,A,Z) &:= \widehat{\E}\left[Y\mid W,A,Z,R_Y=1\right], \\
    \widehat{Q}_\text{CD}(W,A) &:= \int\dd\widehat{P}(Z\mid W,A)\, \widehat{Q}_1(W,A,Z),\\
    \widehat{\psi}_{\text{CD}} &:= \frac{1}{n}\sum_{i=1}^{n}\Delta_a\widehat{Q}_\text{CD}(W_i,a).
\end{align}

\begin{lemma}\label{lemmaCD}
 The CD estimator $\widehat{\psi}_\text{CD}$ is consistent for the ATE given an $s$-admissible pair $(W;Z)$ and under correct specification of $\widehat{Q}_1(W,A,Z)$ and $\widehat{P}(Z\mid W,A)$.
\end{lemma}


In general, this estimator poses difficulties in terms of specification and computation. It requires a consistent probabilistic model $\widehat{P}(Z\mid W,A)$, which can be challenging to specify when $Z$ comprises high-dimensional or mixed-type variables. Employing a parametric approach requires the specification of a number of conditional probability models up to the cardinality of $Z$, depending on the chosen factorization, thereby increasing the chances of model misspecification. Conversely, a fully nonparametric approach, such as kernel density estimation, can affect convergence rates and the behavior of limiting distributions, potentially impeding uniformly valid causal inference \citep{nuisanceModels}.

An alternative approach consists of implementing two \textit{sequential regressions}, with a mean-imputation regression model $\widehat{Q}_1$ defined as in \cref{eqQ0}, followed by a meta-regression model for $Q_2(W,A)={\E}_{Z\mid W,A}\widehat{Q}_1(W,A,Z)$. This strategy involves predicting $\widehat{Q}_1$ for all observations, and then regressing such predicted values against $W$ and $A$ to learn the surface $\widehat{Q}_2$, whose average contrast produces an estimator for the ATE, $\widehat{\psi}_{\text{SR}}$. That is:
\begin{align}\label{eqNR}
    \widehat{Q}_2(W,A) &:= \widehat{\E}_{Z\mid W,A}\widehat{Q}_1(W,A,Z),\\
    \widehat{\psi}_{\text{SR}} &:= \frac{1}{n}\sum_{i=1}^{n}\Delta_a\widehat{Q}_2(W_i,a).
    \label{eqNR2}
\end{align}

\begin{lemma}\label{lemmaSR}
    The SR estimator $\widehat{\psi}_\text{SR}$ is consistent for the ATE given an $s$-admissible pair $(W;Z)$ and under correct specification of $\widehat{Q}_1(W,A,Z)$ and $\widehat{Q}_2(W,A)$.
\end{lemma}

Proofs for \cref{lemmaCD} and \cref{lemmaSR} are provided in \cref{appendixA2,appendixA3} respectively.

The SR estimator holds an advantage over its CD counterpart in that it necessitates the specification of only two regression models, regardless of the cardinality of $Z$. Similar strategies utilizing sequential regressions have been developed for estimation of dynamic treatment effects in longitudinal settings \citep{seqSuper}, for mediation analysis with high-dimensional mediators \citep{ZhengvanderLaan2012,deepmed,highMed}, and for estimating the causal effects of stochastic threshold-based interventions \citep{stochasticThreshold}, among others.

Naïve plug-in estimators, in addition to requiring strict correct specification of all nuisance components for consistency, often fail to achieve an optimal bias-variance tradeoff for the target parameter, as they do not explicitly consider this balance in the estimation process. In contrast, target-aware and multiplyrobust estimators rooted in semiparametric theory offer a more advantageous approach \citep{vanderLaanRubin2006, gruber2009targeted}.


\section{Estimation via TMLE and multiple robustness}
\label{secidTMLE}


Estimation methods based on semiparametric theory are highly valued for their ability to integrate data-adaptive statistical techniques, delivering optimal performance under flexible and realistic assumptions about the underlying causal mechanisms, while maintaining asymptotic guarantees for inference \citep{MLDR}. In typical cases, the derived estimators exhibit \textit{double} or \textit{multiple robustness}, which guarantees consistency even when some particular nuisance components are misspecified \citep{DRbookVariance}. Methodological frameworks for robust semiparametric inference in causal inference and missing data problems include \textit{augmented estimating equations} \citep{DREE,Wei2022}, \textit{M-estimation} \citep{dw}, \textit{semiparametric covariate balancing} \citep{Tang2024}, \textit{debiased machine learning} (DML) \citep{DML,Huber}, and TMLE \citep{TMLEbook1,TMLEbook2}. The latter is particularly prevalent in the assessments of clinical exposures and epidemiological research \citep{Smith2023}, and in settings where informative missingness arise from censoring \citep{rcensoringTMLE,censoringTMLE}.

The TMLE framework pioneered by \citet{vanderLaanRubin2006}, and \citet{gruber2009targeted}, aims at constructing plug-in estimators that are \textit{regular asymptotically linear} (RAL) and efficient. TMLE estimators are built to satisfy the required empirical moment condition of the estimand's \textit{efficient influence function} (EIF), and so they inherit robustness properties from it. The EIF quantifies the estimand's sensitivity to small perturbations in the underlying distribution. Under some boundness and smoothness conditions, it can be derived by employing parametric submodels indexed by a fluctuation parameter \citep{visually}. 

\subsection{The efficient influence function}

Let $P, \widetilde{P} \in \mathfrak{P}$ be probability distributions within a statistical model, and consider a parametric submodel $P_\epsilon = P + \epsilon(\widetilde{P} - P)$ with fluctuation parameter $\epsilon$, defining \textit{smooth paths} within $\mathfrak{P}$ between $P$ and $\widetilde{P}$. For sufficiently small $\epsilon$, these paths can also be expressed in terms of a score function $h=\dv{\epsilon}\log\dv{P_\epsilon}{P}\eval_{\epsilon=0} \in T_P\mathfrak{P}$, as $\dd P_\epsilon = (1 + \epsilon h)\, \dd P$, where $T_P\mathfrak{P} \subseteq L^2_0(P)$ represents the tangent space of the statistical model at $P$, which is embedded within the Hilbert space of $P$-square-integrable functions with zero $P$-mean. An estimand $\Psi[\cdot]$ is said to be \textit{pathwise differentiable} at $P$ if its \textit{Gâteaux derivative}, defined as $\dd\Psi[P;h]:=\dv{}{\epsilon}\Psi[P_\epsilon]\eval_{\epsilon=0}$, exists and has bounded variance for all scores $h$, so it has a unique \textit{Riesz representer} $D^\Psi_P\in T_P\mathfrak{P}$ known as the \textit{canonical gradient} of $\Psi$ at $P$, giving $\dd\Psi[P;h]=\langle {h,D_P^\Psi} \rangle_{L^2_0(P)}$ \citep{van1995efficient,ichimura}. 

An estimator $\widehat{\psi}$ of the parameter $\psi=\Psi[P]$, constructed from i.i.d. data $\{O_i\}_{i=1}^n\sim P$, is RAL with \textit{influence function} $D \in L^2_0(P)$ if it satisfies: \textit{(i)} $\widehat{\psi} - \psi = n^{-1} \sum_{i=1}^n D(O_i) + o_{P}(n^{-1/2})$ and \textit{(ii)} the limiting distribution of $\sqrt{n}\left(\widehat{\psi} - \Psi[P_{\epsilon=1/\sqrt{n}}]\right)$ remains stable under small perturbations of the true distribution for all scores $h$ \citep{bickel1998, van2000}. The second condition, known as \textit{regularity}, is necessary for constructing valid Wald-type confidence intervals via the central limit theorem (CLT). For pathwise differentiable parameters in a fully nonparametric or saturated model, one has that $T_P\mathfrak{P}=L^2_0(P)$. Hence, there is a unique influence function, which is known as the EIF, and it corresponds to the canonical gradient $D_P^\Psi$ of $\Psi$ at $P$ treated as a function of the unit sample $O$. Thus, the EIF can be defined pointwise as $D^\Psi_P: O_i \mapsto \dd\Psi[P; h_{O_i}]$, where $h_{O_i}(O)=\mathbb{I}(O=O_i)\, p(O)^{-1}-1$ corresponds to the score of a point mass fluctuation toward $O_i$ \citep{visually,hines2022demystifying}. The importance of the EIF lies in its ability to capture both the direction of steepest local change in the estimand per unit of information and the \textit{semiparametric efficiency bound}, since $\E [D^\Psi_P(O)^2]$ is the lowest asymptotic variance achievable by RAL estimators of $\psi$ \citep{bickel1998, van2000}. 

\begin{theorem}\label{theoEIF}
    Let $(W;Z)$ be an $s$-admissible pair with $Z\neq\emptyset$ and let $P$ stand for the observed data distribution. Under some technical conditions, the EIF of the $s$-formula for the ATE, $\psi=\Psi[P]$, evaluated at $P$ and observation $O_i\sim P$ is given by: 
\begin{align}\label{eqEIF1}
    D^\Psi_P(O_i) &=  D^Y_P(O_i) +D^Z_P(O_i)+ D^W_P(O_i),\\ \label{eqEIF2}
    D^Y_P(O_i) &:= \frac{A_i-\pi_A(W_i)}{\pi_A(W_i)[1-\pi_A(W_i)]}\frac{R_{Y,i}}{\pi_R(W_i,A_i,Z_i)}\,[Y_i-{Q}_1(W_i,A_i,Z_i)],\\ \label{eqEIF3}
    D^Z_P(O_i) &:= \frac{A_i-\pi_A(W_i)}{\pi_A(W_i)[1-\pi_A(W_i)]}\left[{Q}_1(W_i,A_i,Z_i)- {Q}_2(W_i,A_i) \right],\\ \label{eqEIF4}
    D^W_P(O_i) &:=  \Delta_a {Q}_2(W_i,a) -\psi,
\end{align}

\noindent where $\pi_A(W)=\pr(A=1\mid W)$ is the exposure propensity score, $\pi_R(W,A,Z)=\pr(R_Y=1\mid W,A,Z)$ is the selection propensity score, ${Q}_1(W,A,Z)= {\E}\left[Y\mid W,A,Z,R_Y=1\right]$, and ${Q}_2(W,A)= {\E}_{Z\mid W,A}{Q}_1(W,A,Z)$.
\end{theorem}

A detailed derivation using the point mass contamination approach is provided in \cref{appendixA4}. 


Notably, the EIF is composed of orthogonal components: a double weighted term $D^Y_P$ involving the mean-imputation model $Q_1$, a single weighted term $D^Z_P$ involving both sequential regressions, and an unweighted term involving the meta-regression model $Q_2$. A pivotal property of the EIF is that it satisfies the moment condition $\E D^\Psi_P(O)=0$, which can be leveraged to construct a robust estimating equation \citep{gruber2009targeted}. 

\begin{theorem}\label{theoMR}
    Let $\widetilde{D}$ be the function based on the EIF given in \cref{theoEIF} but replacing $({Q}_1,{Q}_2,{\pi}_A,{\pi}_R)$ by putative models $(\widetilde{Q}_1,\widetilde{Q}_2,\widetilde{\pi}_A,\widetilde{\pi}_R)$. Then, $\E \widetilde{D}(O)=0$ is a valid estimating equation for the ATE $\psi=\Psi[P]$ if at least one of the following conditions is satisfied:
\begin{enumerate}[leftmargin=30pt]
    \item The sequential regression models $\widetilde{Q}_1$ and $\widetilde{Q}_2$ are correctly specified.
    \item The exposure and selection propensity score models $\widetilde{\pi}_A$ and $\widetilde{\pi}_R$ are correctly specified.
    \item The exposure propensity score model $\widetilde{\pi}_A$ and the mean-imputation model $\widetilde{Q}_1$ are correctly specified.
\end{enumerate}
\end{theorem}

A proof is shown in \cref{appendixA5}.

In this context, we formulate a TMLE-based procedure that, under some technical conditions, yields an efficient and RAL plug-in estimator satisfying the empirical version of the EIF moment condition. We call this procedure \textit{targeted sequential regressions} (TSR). 

\subsection{TSR estimator}


The TMLE framework prescribes an iterative procedure designed to optimally update some relevant nuisance components in the estimation task, ensuring that the resulting plug-in estimator satisfies the empirical moment condition of the estimand's EIF. This process can be interpreted as functional gradient descent in the model space, with the constraint that the empirical loss decreases at each step. The procedure involves several key steps: \textit{(i)} initial estimation, \textit{(ii)} construction of the clever covariates, \textit{(iii)} the targeting step, \textit{(iv)} updating of the initial estimates, \textit{(v)} iteration (if needed) and inference \citep{TMLEbook1,TMLEbook2}.

Given an $s$-admissible pair $(W;Z)$, with $Z\neq\emptyset$, the $s$-formula for the ATE $\Psi[P]$ in \cref{eqSformula} depends on the observed data distribution $P$ solely through models $Q_1,Q_2$ when $P_W$ is left fully nonparametrically specified. Consequently, these two sequential parameters are subject to update. We consider the squared error loss function $\ell(y,\widehat{y})=\frac{1}{2}(y-\widehat{y})^2$ along with linear \textit{least favorable path models} with fluctuation parameters $\delta,\gamma$, such that $\delta,\gamma=0$ implies no fluctuation:
\begin{align}
    {Q}_1^{\delta}(W,A,Z) &:= {Q}_1(W,A,Z) + \delta {H}_1(W,A,Z), \text{ under } R_Y=1\\
    {Q}_2^{\gamma}(W,A) &:= {Q}_2(W,A) + \gamma {H}_2(W,A),
\end{align}

\noindent where $H_1,H_2\in L_0^2(P)$ are \textit{clever functions} satisfying:
\begin{align}
    \dv{}{\delta}\ell(Y,Q_1^\delta)\eval_{\delta=0} &= -H_1(Y-Q_1-\delta H_1)\eval_{\delta=0}\, \propto\, H_1(Y-Q_1) = D^Y_P,\\
    \dv{}{\gamma}\ell(Q_1^\delta,Q_2^\gamma)\eval_{\epsilon=\delta=0} &= -H_2(Q_1+\delta H_1-Q_2-\gamma H_2)\eval_{\gamma=\delta=0}\, \propto\, H_2(Q_1-Q_2)=D^Z_P.
\end{align}

Thus, $H_1,H_2$ can be identified in \cref{eqEIF2,eqEIF3}, as follows:
\begin{align}\label{eqH1}
    H_1(W,A,Z) &=\frac{A-\pi_A(W)}{\pi_A(W)[1-\pi_A(W)]}\cdot\frac{1}{\pi_R(W,A,Z)},\\ \label{eqH2}
    H_2(A,W) &=\frac{A-\pi_A(W)}{\pi_A(W)[1-\pi_A(W)]}.
\end{align}

While it has been reported that the negative log-likelihood loss function is more robust to positivity violations and to extreme values in the outputs of the clever functions \citep{ZhengvanderLaan2012}, the squared error loss is more intuitive and natural for unbounded outcomes, and allows the explicit ordinary least square (OLS) solution for the optimal fluctuations.

In this context, given that $\Psi[P]$ is a linear functional in $Q_1,Q_2$ (\cref{appendixA6}), the iterative TMLE procedure can converge in a single step if  initial estimators $\widehat{Q}_1,\widehat{Q}_2$ are sufficiently close to the true values \citep{TMLEbook1,TMLEbook2}. To achieve this, a \textit{super-learning} scheme can be employed. A super-learner comprises a battery of base estimators, a meta-learner or meta-loss, and a cross-validation procedure to generate the best weighted ensemble of the base estimators in the battery \citep{superlearner2007}. Simulation studies have demonstrated the effectiveness of using large sets of flexible base learners, as overfitting the super-learner is hard in practice \citep{superlearners2023}.

\Cref{algo1} presents the complete procedure using basic sample splitting with two halves of the data, $J=J_1\cup J_2$, plus sequential super-learning schemes \citep{seqSuper} and a T-learner for $\widehat{Q}_2$. More involved techniques such as $K$-fold cross-fitting \citep{Huber} and X-learners \citep{kunzel2019metalearners} can also be applied instead. The algorithm outputs the TSR estimator of the ATE $\widehat{\psi}^*$ and its EIF-based asymptotic standard error $\widehat{\sigma}^*$. It can be outlined in the following steps:

\begin{enumerate}[leftmargin=30pt]
    \item Using first half of the data $J_1$ and super-learning schemes, learn models $\widehat{\pi}_{A}$, $\widehat{\pi}_{R}$, along with $\widehat{Q}_1$ in the selected samples ($R_Y=1$). Predict $\widehat{Q}_{1,i}$ for the selected units in $J_1$.
    \item Compute the clever covariates for all $i\in J_1$, $\widehat{H}_{1,i}$ and $\widehat{H}_{2,i}$ according to \cref{eqH1,eqH2}.
    \item Find the optimal fluctuation $\delta^*:=\arg\min \sum_{i\in J_1: R_{Y,i}=1}(Y_i-\widehat{Q}_{1,i}-\delta \widehat{H}_{1,i})^2$.
    \item Let $\widehat{Q}^*_1(W,A,Z):=\widehat{Q}_1(W,A,Z)+\delta^*\widehat{H}_1(W,A,Z)$ and predict $\widehat{Q}^*_{1,i}$ for all $i\in J_1$.
    \item Regress updated predictions $\widehat{Q}_{1,i}^{*}$ against $(W_i,A_i)$ using super-learning schemes to build initial surfaces $\widehat{Q}_2$ separately for each treatment arm. Predict $\widehat{Q}_{2,i}$ for all $i\in J_1$.
    \item Find the optimal fluctuation $\gamma^*:=\arg\min \sum_{i\in J_1}(\widehat{Q}_{1,i}^{*}-\widehat{Q}_{2,i}-\gamma \widehat{H}_{2,i})^2$.
    \item Let $\widehat{Q}^*_2(W,A):=\widehat{Q}_2(W,A)+\gamma^*\widehat{H}_2(W,A)$.
    \item Using the second half of the data, plug in the updated nuisance parameters and the empirical distribution $\widehat{P}_W$ into the $s$-formula $\Psi[\cdot]$ to produce an estimate  $\widehat{\psi}^*$. Compute values $D_i^*$ of the EIF for every $i\in J_2$. The asymptotic variance of $\widehat{\psi}^*$ can be computed as the empirical variance of $D_i^*$ divided by the sample size in $J_2$.
\end{enumerate} 

\begin{algorithm}[t]
\caption{TSR estimating procedure for the ATE}
\begin{algorithmic}[1]\label{algo1}
\Require data ($n_1$ samples in $J_1$ and $n_2$ samples in $J_2$), super-learning schemes 
\Ensure TSR estimate of the ATE $\widehat{\psi}^*$, and asymptotic standard error $\widehat{\sigma}^*$

\State Learn $\widehat{Q}_1(W,A,Z)$ using super-learner with data in $\{i\in J_1 : R_{Y,i}=1\}$ \Comment{step (i)}

\State \phantom{Learn} $\widehat{Q}_{1,i}\gets\widehat{Q}_1(W_i,A_i,Z_i)$ for $i\in J_1 : R_{Y,i}=1$

\State Learn $\widehat{\pi}_A(W)$, $\widehat{\pi}_R(W,A,Z)$ using super-learners with data in $J_1$ 

\State Let $\widehat{H}_2:(W,A)\mapsto \dfrac{A-\widehat{\pi}_A(W)}{\widehat{\pi}_A(W)[1-\widehat{\pi}_A(W)]}$ and $\widehat{H}_{2,i}\gets\widehat{H}_2(W_i,A_i)$ for $i\in J_1$ \Comment{step (ii)}

\State \phantom{Let} $\widehat{H}_1:(W,A,Z)\mapsto \dfrac{\widehat{H}_2(W,A)}{\widehat{\pi}_R(W,A,Z)}$ and $\widehat{H}_{1,i}\gets\widehat{H}_1(W_i,A_i,Z_i)$ for $i\in J_1$

\State Let $\delta^*\gets\dfrac{\sum_{i\in J_1:R_{Y,i}=1}\widehat{H}_{1,i}(Y_i-\widehat{Q}_{1,i})}{\sum_{i\in J_1:R_{Y,i}=1}(\widehat{H}_{1,i})^2}$ \Comment{step (iii)}

\State Update $\widehat{Q}_1^{*}:(W,A,Z) \mapsto \widehat{Q}_1(W,A,Z)+ {\delta}^*\widehat{H}_1(W,A,Z)$ \Comment{step (iv)}

\State \phantom{Update} $\widehat{Q}_{1,i}^{*} \gets \widehat{Q}^*_1(W_i,A_i,Z_i)$  for $i\in J_1$

\For{$a \in \{0,1\}$}
    \State Learn $\widehat{Q}_2(W,a)\gets\widehat{\E}\left[\widehat{Q}_{1,i}^{*}\mid W,A=a\right]$ using super-learner with data in $\{i\in J_1 : A_i=a\}$ \Comment{step (v)}
    \State \phantom{Learn} $\widehat{Q}_{2,i}\gets \widehat{Q}_2(W_i,a)$ for $i\in J_1 : A_i=a$ 
\EndFor

    \State Let $\gamma^*\gets\dfrac{\sum_{i\in J_1}\widehat{H}_{2,i}(\widehat{Q}_{1,i}^{*}-\widehat{Q}_{2,i})}{\sum_{i\in J_1}(\widehat{H}_{2,i})^2}$ \Comment{step (vi)}
    
\State Update $\widehat{Q}_2^{*}:(W,A) \mapsto \widehat{Q}_2(W,A)+ {\gamma}^*\widehat{H}_2(W,A)$ \Comment{step (vii)}

\State Let $\widehat{\psi}^*\gets {n_2}^{-1}\sum_{i\in J_2}\Delta_a Q_2^{*}(W_i,a)$ \Comment{step (viii)}

\State \phantom{Let} ${D}^*_i\gets \widehat{H}_{1}(W_i,A_i,Z_i)\,R_{Y,i}\,[Y_i-\widehat{Q}^*_{1}(W_i,A_i,Z_i)]$ 

\State \phantom{Let XXxx} $+\widehat{H}_{2}(W_i,A_i)\,[\widehat{Q}^*_{1}(W_i,A_i,Z_i)-\widehat{Q}^*_{2}(W_i,A_i)] +\Delta_a \widehat{Q}^*_{2}(W_i,a) - \widehat{\psi}^*$ for $i\in J_2$

\State \phantom{Let} $\widehat{\sigma}^*\gets\left[\widehat{\text{var}}\left(\{D^*_i\}_{i\in J_2}\right)/n_2\right]^{1/2}$ 

\State\Return $\widehat{\psi}^*,\widehat{\sigma}^*$
\end{algorithmic}
\end{algorithm}


In contrast to the DIPW estimator, and to the naïve plug-in estimators discussed in \cref{secPlugin}, the TSR estimator is multiplyrobust, inheriting all the robustness properties from the EIF. Specifically, it remains consistent under the same conditions outlined in \cref{theoMR}. Condition \textit{(i)} is necessary to ensure the consistency of the naïve plug-in sequential regression estimator, whereas condition \textit{(ii)} is critical for the consistency of the DIPW estimator. The TSR estimator offers an additional layer of robustness under condition \textit{(iii)}, for which it leverages the exposure propensity score model and the mean-imputation model. Interestingly, this result finds rationale in the classical literature of recovery from missing data via imputations. Specifically, if $\widehat{Q}_1(W_i,A_i,Z_i)$ consistently estimates the expected conditional outcome, then $Y_i$ can be replaced by it when observed and imputed by it when missing. Hence, a valid single weighted estimator of the ATE can be constructed by treating $\widehat{Q}_1(W_i,A_i,Z_i)$ as an imputed \textit{pseudo-outcome}, and employing IPW with weights given solely by the exposure propensity score model. In a sense, the TSR estimator capitalizes on the collective robustness conditions of regression-based, IPW-based, and imputation-based estimators.

Besides, if $P^*$ is any observed data distribution consistent with $(P_W,Q_1^*,Q_2^*)$, the TMLE estimator $\widehat{\psi}^*=\Psi[P^*]$ is RAL under certain technical conditions related to two components of the asymptotic bias: \textit{(i)} the second-order remainder $\widehat{\psi}^*-\psi+\E D^\Psi_{P^*}(O)$, and \textit{(ii)}
the empirical process term $n^{-1}\sum_{i=1}^n\left[D^\Psi_{P*}(O_i)-D^\Psi_P(O_i) \right]-\E\left[D^\Psi_{P*}(O)-D^\Psi_P(O)\right]$. Both should vanish at rate $o_P(n^{-1/2})$. There are two methodological approaches to address the second requirement. The first is to impose the Donsker class condition, which requires that the estimators employed exhibit bounded complexity, preventing rapid entropy increases with the sample size \citep{van1996weak}. This ensures that the empirical process of the EIF converges weakly to a Gaussian process, enabling standard asymptotic inference. However, this condition limits the types of algorithms that can be employed for learning, excluding many that are highly adaptive to the data and prone to overfitting \citep{DML}. Another approach relaxes the Donsker condition, allowing the use of data-adaptive estimators, but at the cost of decreased sample size through sample-splitting, as implemented by default in \cref{algo1}. Finally, to address the requirement placed upon the second-order remainder, certain products of the nuisance parameters estimators must converge at specific rates. For this, it is sufficient that the \textit{models dealing with confounding}, $\widehat{\pi}_A$ and $\widehat{Q}_2$, are consistent and their bias vanish at rate $o_P(n^{-1/4})$, while the \textit{models dealing with selection}, $\widehat{\pi}_R$ and $\widehat{Q}_1$, are consistent and converge as fast as $o_P(n^{-1/8})$ (see \cref{appendixA7}).



\section{Simulations}
\label{secSimulation}


In this section, we assess the performance of the proposed TSR estimator in comparison to other alternatives across two distinct benchmark data-generating processes. These simulation setups aim to illustrate some of the theoretical results, examine the behavior of estimators in finite data, and contrast them based on bias, mean squared error (MSE), and coverage probability of their associated confidence intervals. 

\subsection{Simulation setup I}

\begin{figure}[t]
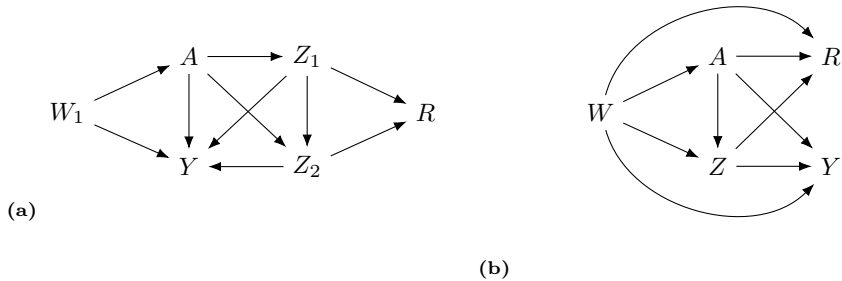

\centering
\begin{subfigure}[c]{.4\textwidth}
  \centering
\tikz[scale=0.8, transform shape]{
    \node (w) {$W_1$};
    \node (a) [right = of w, yshift=0.75cm] {$A$};
    \node (m) [right = of a] {$Z_1$};
    \node (y) [below = of a] {$Y$};
    \node (z) [below = of m, yshift=+0.05cm] {$Z_2$};
    \node (r) [right = of m, yshift=-0.75cm] {$R_Y$};
    
    \path (w) edge[style=directed] (a);
    \path (w) edge[style=directed] (y);
    \path (a) edge[style=directed] (z);
    \path (a) edge[style=directed] (m);
    \path (a) edge[style=directed] (y);
    \path (z) edge[style=directed] (y);
    \path (m) edge[style=directed] (z);
    \path (m) edge[style=directed] (y);
    \path (m) edge[style=directed] (r);
    \path (z) edge[style=directed] (r);
}
    \caption{}
    \label{fig3a}
\end{subfigure}%
\begin{subfigure}[c]{.4\textwidth}
  \centering
\tikz[scale=0.8, transform shape]{
    \node (h) {$W$};
    \node (a) [right = of h, yshift=0.75cm] {$A$};
    \node (s) [right = of a] {$R_Y$};
    \node (m) [below = of a] {$Z$};
    \node (y) [below = of s] {$Y$};

    \path (a) edge[style=directed] (y);
    \path (h) edge[style=directed] (a);
    \path (a) edge[style=directed] (m);
    \path (h) edge[style=directed] (m);
    \path (m) edge[style=directed] (y);
    \path (m) edge[style=directed] (s);
    \path (a) edge[style=directed] (s);
    \path (h) edge[style=directed,bend right=60] (y);
    \path (h) edge[style=directed,bend left=60] (s);
}
    \caption{}
    \label{fig3b}
\end{subfigure}%
\caption{$m$-graph used in (a) simulation setup I and (b) application case.}
\label{fig3}
\end{figure}

We generate 5\,000 i.i.d. samples from an SCM with associated $m$-graph presented in \cref{fig3a}. Details regarding the specification of exogenous noises and causal mechanisms can be found in \cref{appendixA8}. In this setup, $W_1$ serves as a confounder, covariates $Z_1,Z_2$ act as mediators of the effect of the point exposure $A$ upon the outcome $Y$, and $(\{W_1\};\{Z_1,Z_2\})$ is the only minimal $s$-admissible pair for recovering the ATE. 

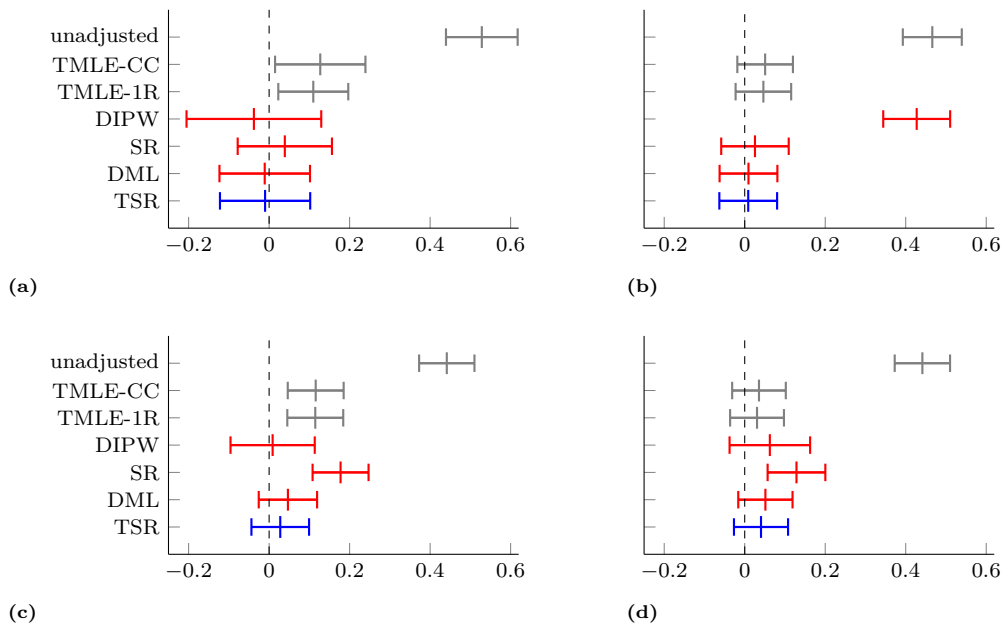
\begin{figure}[t]
\centering
\begin{subfigure}{.475\textwidth}
    \centering
    \begin{tikzpicture}
    \begin{axis}[
    scale=0.85,
    height=5.0cm,
    width=7.0cm,  
      xmax=0.62,
      xmin=-0.25,
      ymin=0,
      ymax=8,
      ytick={1,...,7},
      xtick={-0.2,0,0.2,0.4,0.6},
      axis y line*=left,
      axis x line*=bottom,
      yticklabels={TSR, DML, SR, DIPW, TMLE-1R, TMLE-CC, unadjusted}]
    \addplot+ [blue, solid, style=thick, boxplot prepared={
        lower whisker=-0.12213, median=-0.01002, upper whisker=0.10208
        }] coordinates {}; 
    \addplot+ [red, solid, style=thick, boxplot prepared={
        lower whisker=-0.12339, median=-0.01091, upper whisker=0.10158
        }] coordinates {}; 
    \addplot+ [red, solid, style=thick, boxplot prepared={
        lower whisker=-0.07811, median=0.039, upper whisker=0.15611
        }] coordinates {};
    \addplot+ [red, solid, style=thick, boxplot prepared={
        lower whisker=-0.2052, median=-0.03784, upper whisker=0.12951
        }] coordinates {};
    \addplot+ [gray, solid, style=thick, boxplot prepared={
        lower whisker=0.0228, median=0.10972, upper whisker=0.19664
        }] coordinates {};
    \addplot+ [gray, solid, style=thick, boxplot prepared={
        lower whisker=0.01481, median=0.12704, upper whisker=0.23928
        }] coordinates {};
    \addplot+ [gray, solid, style=thick, boxplot prepared={
        lower whisker=0.43938, median=0.52842 , upper whisker=0.61747
        }] coordinates {};
    \addplot[color=black,dashed] coordinates {
        		(0,8)
        		(0,0)
        	};
    \end{axis}
\end{tikzpicture}%
    \caption{}
    \label{fig4a}
\end{subfigure}%
\hfill
\begin{subfigure}{.475\textwidth}
    \begin{tikzpicture}
    \begin{axis}[
    scale=0.85,
    height=5.0cm,
    width=7.0cm,   
      xmax=0.62,
      xmin=-0.25,
      ymin=0,
      ymax=8,
      ytick={1,...,7},
       xtick={-0.2,0,0.2,0.4,0.6},
      axis y line*=left,
      axis x line*=bottom,
      yticklabels={,,}]
    \addplot+ [blue, solid, style=thick, boxplot prepared={
        lower whisker=-0.06303, median=0.00873, upper whisker=0.08048
        }] coordinates {}; 
    \addplot+ [red, solid, style=thick, boxplot prepared={
        lower whisker=-0.06242, median=0.00936, upper whisker=0.08115
        }] coordinates {}; 
    \addplot+ [red, solid, style=thick, boxplot prepared={
        lower whisker=-0.05838, median=0.02548, upper whisker=0.10935
        }] coordinates {};
    \addplot+ [red, solid, style=thick, boxplot prepared={
        lower whisker=0.34415, median=0.42729, upper whisker=0.51042
        }] coordinates {};
    \addplot+ [gray, solid, style=thick, boxplot prepared={
        lower whisker=-0.02263, median=0.04639, upper whisker=0.11541
        }] coordinates {};
    \addplot+ [gray, solid, style=thick, boxplot prepared={
        lower whisker=-0.01835, median=0.05073, upper whisker=0.1198
        }] coordinates {};
    \addplot+ [gray, solid, style=thick, boxplot prepared={
        lower whisker=0.39282, median=0.46607, upper whisker=0.53931
        }] coordinates {};
    \addplot[color=black,dashed] coordinates {
        		(0,9)
        		(0,0)
        	};
    \end{axis}
\end{tikzpicture}%
    \caption{}
    \label{fig4b}
\end{subfigure}%
\vskip\baselineskip
\begin{subfigure}{.475\textwidth}
    \centering
    \begin{tikzpicture}
    \begin{axis}[
    scale=0.85,
    height=5.0cm,
    width=7.0cm,  
      xmax=0.62,
      xmin=-0.25,
      ymin=0,
      ymax=8,
      ytick={1,...,7},
      xtick={-0.2,0,0.2,0.4,0.6},
      axis y line*=left,
      axis y line*=left,
      axis x line*=bottom,
      yticklabels={TSR, DML, SR, DIPW, TMLE-1R, TMLE-CC, unadjusted}]
    \addplot+ [blue, solid, style=thick, boxplot prepared={
        lower whisker=-0.04394, median=0.02765, upper whisker=0.09924
        }] coordinates {}; 
    \addplot+ [red, solid, style=thick, boxplot prepared={
        lower whisker=-0.02579, median=0.0467, upper whisker=0.1192
        }] coordinates {}; 
    \addplot+ [red, solid, style=thick, boxplot prepared={
        lower whisker=0.10801, median=0.17753, upper whisker=0.24705
        }] coordinates {};
    \addplot+ [red, solid, style=thick, boxplot prepared={
        lower whisker=-0.0959, median=0.00874, upper whisker=0.11337
        }] coordinates {};
    \addplot+ [gray, solid, style=thick, boxplot prepared={
        lower whisker=0.04518, median=0.11467, upper whisker=0.18416
        }] coordinates {};
    \addplot+ [gray, solid, style=thick, boxplot prepared={
        lower whisker=0.04628, median=0.11572, upper whisker=0.18516
        }] coordinates {};
    \addplot+ [gray, solid, style=thick, boxplot prepared={
        lower whisker=0.37264, median=0.4414, upper whisker=0.51015
        }] coordinates {};
    \addplot[color=black,dashed] coordinates {
        		(0,9)
        		(0,0)
        	};
    \end{axis}
\end{tikzpicture}%
    \caption{}
    \label{fig4c}
\end{subfigure}%
\hfill
\begin{subfigure}{.475\textwidth}
    \begin{tikzpicture}
    \begin{axis}[
    scale=0.85,
    height=5.0cm,
    width=7.0cm,   
      xmax=0.62,
      xmin=-0.25,
      ymin=0,
      ymax=8,
      ytick={1,...,7},
      xtick={-0.2,0,0.2,0.4,0.6},
      axis y line*=left,
      axis x line*=bottom,
      yticklabels={,,}]
    \addplot+ [blue, solid, style=thick, boxplot prepared={
        lower whisker=-0.02671, median=0.0405, upper whisker=0.10771
        }] coordinates {}; 
    \addplot+ [red, solid, style=thick, boxplot prepared={
        lower whisker=-0.01601, median=0.05141, upper whisker=0.11883
        }] coordinates {}; 
    \addplot+ [red, solid, style=thick, boxplot prepared={
        lower whisker=0.05693, median=0.12862, upper whisker=0.20032
        }] coordinates {};
    \addplot+ [red, solid, style=thick, boxplot prepared={
        lower whisker=-0.03763, median=0.06243, upper whisker=0.16249
        }] coordinates {};
    \addplot+ [gray, solid, style=thick, boxplot prepared={
        lower whisker=-0.03628, median=0.0306, upper whisker=0.09749
        }] coordinates {};
    \addplot+ [gray, solid, style=thick, boxplot prepared={
        lower whisker=-0.03116, median=0.03546, upper whisker=0.10209
        }] coordinates {};
    \addplot+ [gray, solid, style=thick, boxplot prepared={
        lower whisker=0.37264, median=0.4414, upper whisker=0.51015
        }] coordinates {};
    \addplot[color=black,dashed] coordinates {
        		(0,9)
        		(0,0)
        	};
    \end{axis}
\end{tikzpicture}%
    \caption{}
    \label{fig4d}
\end{subfigure}%
\caption{Bias from estimators of the ATE with 95\% confidence intervals: (a) 50\% missingness rate, (b) 25\% missingness rate and model misspecification for $\widehat{\pi}_{A}$, (c) 15\% missingness rate and model misspecification for $\widehat{Q}_1$, and (d) 15\% missingness rate and model misspecification for $\widehat{Q}_2$ and $\widehat{\pi}_{R}$. In \textcolor{blue}{blue}: proposed TSR estimator; in \textcolor{red}{red}: alternative consistent estimators; in \textcolor{gray}{gray}: inconsistent estimators. Values are in Cohen's $d$ scale.}
\label{fig4}
\end{figure}

We compare the TSR estimator against three admissible alternatives: the DIPW estimator, in \cref{eqIPW1}, the naïve plug-in sequential regression (SR) estimator, in \cref{eqNR2}, and the DML-based estimator from \citet{Huber}. In addition, we include three estimators that are not admissible in this context, and therefore inconsistent: the unadjusted estimator, the TMLE estimator applied to complete cases (TMLE-CC) ---without accounting for the selection propensity score model---, and the TMLE estimator using a single regression adjustment (TMLE-1R), with a selection propensity score model based solely on pre-exposure variables $W=\{W_1\}$. The TMLE-CC estimator is included to quantify the magnitude of selection bias inherent in complete-case analysis, while TMLE-1R aids in quantifying the residual bias due to insufficient adjustment, as the post-exposure variables $Z=\{Z_1,Z_2\}$ are needed to $d$-separate the outcome from its selection indicator, but cannot be included in single-regression approaches due to their mediator status. These inconsistent estimators are shown only in \cref{fig4} to highlight the potentially erroneous inferences from complete-case analysis and pre-exposure-only methods, which are commonly used approaches for addressing missing data in applied studies.

We assess the performance and robustness of admissible estimators across four different scenarios:
\begin{enumerate}[leftmargin=30pt,label=(\alph*)]
    \item High missingness rate: missing 50\% of the outcome observations, and no model misspecifications,
    \item Moderate missingness rate, 25\%, and model misspecification for $\widehat{\pi}_{A}$,
    \item Moderate-low missingness rate, 15\%, and model misspecification for $\widehat{Q}_1$,
    \item Moderate-low missingness rate, 15\%, and model misspecification for $\widehat{Q}_2, \widehat{\pi}_R$.
\end{enumerate}

Whenever models are not purposely misspecified, super-learning schemes are employed to estimate all nuisance parameters. The battery of base learners includes sample averages, generalized linear models, and multivariate adaptive regression splines. Stacking weights are determined through cross-validation, using the default scheme provided by the \texttt{sl3} package in {R}. No sample splitting is performed, as these algorithms can be sufficiently regularized. The standard error for the DIPW estimator is calculated using a robust sandwich variance estimator, while the confidence interval for the SR estimator is derived through bootstrapping. Inference results from 200 repeated simulations, showing the average estimates and 95\% confidence intervals are presented in \cref{fig4}. \Cref{tab1} provides a summary of the admissible estimators' performance in terms of bias, MSE, and the coverage probability of the constructed confidence intervals.

Scenario (a) features strong selection bias, reflected in the deviation of the TMLE-CC bias estimate in \cref{fig4a}. As expected here, the TMLE-1R estimator fails to recover the ATE due to the lack of $s$-admissibility of the pair $(\{W_1\},\emptyset)$. The DML-based and TSR estimators exhibit nearly identical results in terms of bias, MSE, and coverage probability. Although their estimated biases are the lowest, their coverage probability falls below the nominal 95\%. This shortfall is attributed to EIF-based estimators of the asymptotic variance often underestimating the true asymptotic uncertainty in finite samples \citep{IFVariance}, a phenomenon that may be amplified in scenarios where predictions from regression models are used as observed pseudo-outcomes in sequential regression models, as is our case.

Scenarios (b) to (d) involve intentionally misspecified models for certain components of the nuisance parameters. Since EIF-based estimators of the asymptotic variance are generally consistent only when all relevant nuisance components are correctly specified \citep{DRbookVariance,shook2024Variance}, it is anticipated that the coverage of most estimators will be inadequate under these misspecified conditions. In case (b), the exposure propensity score model is misspecified, resulting in biased estimates of the DIPW estimator. The DML-based and TSR estimators remain unaffected by this misspecification (\cref{fig4b}). In case (c), the mean-imputation model is misspecified, severely affecting the SR estimator, even when only 15\% of the outcome observations are missing. Here, the DIPW estimator outperforms others in terms of bias and coverage, as it is unaffected by this particular misspecification (\cref{fig4c}). The coverage probability of the DML-based and TSR estimators is reduced, with TSR being more robust. This is because in the TSR procedure, predictions from the $Q_1$ model undergo two updates using well-specified propensity scores, smoothing out misspecification-induced deviations towards the DIPW estimate \citep{MLDR}. Finally, scenario (d) involves simultaneous misspecifications in the selection propensity score and the meta-regression model, affecting all estimators (\cref{fig4d}). In this case, the TSR estimator performs best in terms of bias and MSE, with its coverage probability being more robust than that of the DML-based estimator, owing to the $Q_2$ update using well-specified exposure propensity scores.

\begin{table}[t]
\caption{Performance of admissible estimators of the ATE for setup I in relation to bias, MSE, and coverage of 95\% confidence intervals across four scenarios: (a) 50\% missingness rate, (b) 25\% missingness rate and model misspecification for $\widehat{\pi}_{A}$, (c) 15\% missingness rate and model misspecification for $\widehat{Q}_1$, and (d) 15\% missingness rate and model misspecification for $\widehat{Q}_2$ and $\widehat{\pi}_{R}$.}
\centering
\begin{tabular}{lrrrrrrrrrrrr}
\toprule
& \multicolumn{3}{c}{Scenario (a)} & \multicolumn{3}{c}{Scenario (b)} & \multicolumn{3}{c}{Scenario (c)} & \multicolumn{3}{c}{Scenario (d)}\\ 
Estimator & Bias & MSE & Cover. & Bias & MSE & Cover. & Bias & MSE & Cover. & Bias & MSE & Cover.\\ \midrule
DIPW & -0.27 & 0.64 & 94.5 & 3.00 & 3.01 & 0.0 & 0.06 & 0.36 & 94.5 & 0.38 & 0.45 & 84.5 \\
SR   & 0.27 & 0.50 & 92.5 & 0.18 & 0.35 & 94.0 & 1.24 & 1.27 & 0.0 & 0.90 & 0.94 & 6.0\\
DML   & -0.08 & 0.44 & 92.0 & 0.07 & 0.29 & 90.5 & 0.33 & 0.42 & 75.0 & 0.36 & 0.44 & 64.0 \\
TSR & -0.10 & 0.46 & 91.5 & 0.06 & 0.30 & 90.5 & 0.19 & 0.33 & 89.5 & 0.28 & 0.38 & 77.5  \\ \bottomrule
\end{tabular}
\label{tab1}
\end{table}

\subsection{Simulation setup II}

This setup illustrates the relative performance by the TSR estimator using two different minimal $s$-admissible pairs. We simulate data from an SCM with the semi-Markovian $m$-graph depicted in \cref{fig2a} in \cref{secSAC}, with detailed specifications provided in \cref{appendixA9}. This $m$-graph features two minimal $s$-admissible pairs: $(\{B_1,C_1,C_2\};\emptyset)$ and $(\{B_1\};\{C_2\})$. By using the first pair, we compute two estimators: the naïve plug-in regression estimator and the TMLE-1R estimator, both adjusting for $W=\{B_1,C_1,C_2\}$ only. Despite $W$ including descendants of the exposure $A$, it satisfies the SAC. Additionally, we compute \citet{Huber}'s DML-based estimator, along with the TSR estimator from \cref{algo1}, both utilizing the minimal $s$-admissible pair $(\{B_1\};\{C_2\})$. These last two estimators are compared across two scenarios with a moderate missingness rate of 25\% missing outcome data: \textit{(a')} using super-learning specifications for nuisance function estimation, as in the previous setup, and  \textit{(b')} with a misspecified meta-regression model. Results from 200 simulations of 5\,000 i.i.d. samples from the SCM, showing average bias, MSE, and 95\% confidence interval coverage probability, are presented in \cref{tab2}.

\begin{table}[t]
\caption{Performance of admissible estimators of the ATE for setup II in relation to bias, MSE and coverage of 95\% confidence intervals across two scenarios with 25\% of outcome data missing: (a) no misspecification, and (b) model misspecification for $\widehat{Q}_2$}
\centering
\begin{tabular}{lcrrrrrr}
\toprule
& & \multicolumn{3}{c}{Scenario (a')} & \multicolumn{3}{c}{Scenario (b')} \\ 
Estimator & Adj. Pair. & Bias & MSE & Cover. & Bias & MSE & Cover.\\ \midrule
Plug-in reg. & $(\{B_1,C_1,C_2\};\emptyset)$ & -0.12 & 0.33 & 81.0 &  &  &   \\
TMLE-1R  & $(\{B_1,C_1,C_2\};\emptyset)$ & -0.12 & 0.36 & 86.5 &  & &  \\
DML & $(\{B_1\};\{C_2\})$                & -0.01 & 0.31 & 92.0 & 0.44 & 0.53 & 63.0 \\
TSR & $(\{B_1\};\{C_2\})$                & 0.01 & 0.31 & 91.5 & 0.36 & 0.47 & 73.5  \\ \bottomrule
\end{tabular}
\label{tab2}
\end{table}

In this setup, estimators utilizing the minimal $s$-admissible pair $(\{B_1\};\{C_2\})$ outperform the others in scenario \textit{(a')} in terms of bias, MSE, and coverage, despite requiring one additional regression model. However, their performance deteriorates in scenario \textit{(b')} relative to single-regression estimators, as the latter do not rely on the misspecified model $\widehat{Q}_2$. Specifically, the TMLE-1R approach provides a doubly-robust estimator, achieving consistency when either $\widehat{Q}_1(W,A)$ is correctly specified or both $\widehat{\pi}_A(W)$ and $\widehat{\pi}_R(W,A)$ are correctly specified. Moreover, the TSR estimator demonstrates a relatively greater robustness to misspecification compared to the DML-based estimator. This highlights the intricate trade-offs and interplay between graph-based adjustment criteria and the robustness properties of derived estimators. 


\section{Applications}
\label{secApplication}

In this section, we apply the developed procedures in the problem of estimating the causal effects of pharmacological treatment for ADHD upon the scores obtained by diagnosed Norwegian schoolchildren in national tests using observational data.

ADHD is a neurodevelopmental disorder identified by a consistent pattern of inattention and hyperactivity-impulsivity that impacts an individual's social, academic, or occupational functioning \citep{icd11, apa2022dsmvtr}. Roughly 2---6\% of children and adolescents receive a diagnosis of ADHD worldwide, making it among the most prevalent mental health conditions in young individuals \citep{cortese2023incidence}. Those diagnosed with ADHD often experience an elevated occurrence of various negative life outcomes, such as reduced quality of life, substance abuse, accidental injuries, academic underachievement, and unemployment \citep{faraone2021world}. Children's school performance has been examined due to its potential as an early indicator of other long-term adverse outcomes \citep{shaw2012systematic, fredriksen2014childhood}. While treatment with stimulant medication has documented effects on the reduction of symptoms \citep{Cortese2018-qd}, its impact is more modest on various school-related outcomes including educational performance, classroom behavior, grade point averages, and school completion rates \citep{faraone2021world, Storebo2015dz}. Yet, it has been noted that these small benefits may not fully translate into improved learning or better standardized test scores \citep{pelham2022effect,JANGMO2019423}. Given such weak and ambiguous evidence, the clinical significance of these effects remains unclear, highlighting the need for accurate and unbiased estimations of effect sizes.

In the Norwegian educational system, compulsory national tests assessing basic skills in numeracy, reading (Norwegian), and English are administered during grades 5, 8, and 9. Over the recent years, the average participation rate for all subjects in the grade 8 national test has been around 92\% \citep{udir2022eigthninth}. Exemption and abstentions may be driven by endogenous factors, potentially introducing selection bias. Specifically, students with poorer performance might be anticipated to have higher rates of missing test scores. For instance, most exemption cases are granted to pupils already receiving supplementary special education \citep{oslo2010exemption}. The selection process could be influenced by the exposure and its consequences as well, given that methylphenidate is the most commonly prescribed pharmacological treatment for ADHD and that sleep problems and weight loss are frequently reported adverse effect \citep{graham2008adverse,Storebo2015dz}. School change, in the transition from primary to lower secondary school, along with other indicators of post-exposure health status, might be relevant factors to consider as well.

We assess the impact of pharmacological treatment with stimulant medication upon the numeracy and reading (Norwegian) test scores at grade 8 obtained by Norwegian children diagnosed with ADHD. By integrating information from national registries, we compile data on the medication history and national test scores of all children diagnosed with ADHD born between 2000 and 2007 in Norway, who would go to take the national test up to 2021. We exclude those with severe comorbid disorders and those with missing test scores at grade 5 (totaling $9\,352$ individuals). Variables at the student, family, and school levels are linked from the Norwegian Prescription Database (NorPD), the Norwegian Patient Registry (NPR), the Database for Control and Payment of Health Reimbursement (KUHR), Statistics Norway (SSB), and the Medical Birth Registry of Norway (MBRN). We leverage data on students' and parents' diagnoses and their consultations with medical services during pre-exposure and post-exposure periods. Indicators of post-exposure medical status serve as proxies for adverse effects of treatment and its consequences. To operationalize relevant variables, we employ the following grouping:
\begin{itemize}
    \item \textbf{pre-exposure covariates} $W$: Sex at birth, birth year/month cohorts, birth parity number, raw scores at grade 5 national test for numeracy and reading, missingness indicator for scores at grade 5 English national test, mother's education level, mother's age at birth, student's and parents' diagnoses and medical consultations for related comorbid disorders, school identification (fixed effect), prior dispensations of ADHD stimulant medication for at least 90 days, and duration of prior treatment.
    \item \textbf{Exposure} $A$: 
    Having received dispensations of ADHD stimulant medication for at least 75\% of the prescribed treatment period between the start of grade 6 and the national test in grade 8.
    \item \textbf{post-exposure covariates} $Z$: Student's diagnoses and medical consultations for related comorbid disorders, school change in transition from primary to lower secondary school.
    \item \textbf{Outcomes} $Y$: Raw scores at grade 8 national test for numeracy and reading.
\end{itemize}

The $m$-graph in \cref{fig3b} represents the clinical and background causal knowledge of the problem. The missingness rate on test scores at grade 8 is about 8\% for both subjects. This rate is not expected to induce a disproportionate amount of selection bias but, from a public health standpoint, it could still lead to shifts in conclusions and policy design that may not be fully aligned with the target population. 

\begin{figure}[t]
\centering
\begin{subfigure}{.475\textwidth}
    \centering
    \begin{tikzpicture}
    \begin{axis}[
    scale=0.85,
    height=5.0cm,
    width=7.0cm,   
      xmax=1.8,
      xmin=0.0,
      ymin=0,
      ymax=8,
      ytick={1,...,7},
      axis y line*=left,
      axis x line*=bottom,
      yticklabels={TSR, DML, SR, DIPW, TMLE-1R ,TMLE-CC, Unadjusted},
      xlabel={raw score points}]
    \addplot+ [blue, solid, style=thick, boxplot prepared={
        lower whisker=0.6917787799836841, median=1.0637324279991798, upper whisker=1.4356860760146755
        }] coordinates {}; 

    \addplot+ [red, solid, style=thick, boxplot prepared={
        lower whisker=0.60749735198, median=0.979451, upper whisker=1.35140464802
        }] coordinates {}; 
    
    \addplot+ [red, solid, style=thick, boxplot prepared={
        lower whisker=0.4460518064915133, median=0.73333648707, upper whisker=1.0206211676495713
        }] coordinates {}; 
    
    \addplot+ [red, solid, style=thick, boxplot prepared={
        lower whisker=0.3828558496113987, median=0.8417471677846996, upper whisker=1.3006384859580004 
        }] coordinates {};

    \addplot+ [gray, solid, style=thick, boxplot prepared={
        lower whisker=0.8609987432665009, median=1.1663839728113374, upper whisker=1.471769202356174
        }] coordinates {}; 
        
    \addplot+ [gray, solid, style=thick, boxplot prepared={
        lower whisker=1.06959477046, median=1.37498, upper whisker=1.68036522954
        }] coordinates {}; 
    
    \addplot+ [gray, solid, style=thick, boxplot prepared={
        lower whisker=0.2368527, median=0.6614411, upper whisker=1.0860295
        }] coordinates {}; 
    
    \addplot[color=black,dashed] coordinates {
        		(0,8)
        		(0,0)
        	};
    \end{axis}
    \begin{axis}[
    scale=0.85,
    height=5.0cm,
    width=7.0cm,     
      xmax=1.8/11.1,
      xmin=0.0/11.1,
      ymin=0,
      ymax=8,
      ytick={1,...,7},
      axis y line*=left,
      axis x line*=top,
      yticklabels={,,},
      xticklabel style={/pgf/number format/fixed},
      xlabel={Cohen's $d$}]
      \end{axis}
\end{tikzpicture}%
    \caption{}
    \label{fig5a}
\end{subfigure}%
\hfill
\begin{subfigure}{.475\textwidth}
    \begin{tikzpicture}
    \begin{axis}[
    scale=0.85,
    height=5.0cm,
    width=7.0cm,  
      xmax=1.2,
      xmin=-0.3,
      ymin=0,
      ymax=8,
      ytick={1,...,7},
      axis y line*=left,
      axis x line*=bottom,
      yticklabels={,,},
      xlabel={raw score points}]
    \addplot+ [blue, solid, style=thick, boxplot prepared={
        lower whisker=0.3302858222891309, median=0.6614349946501556, upper whisker=0.9925841670111804
        }] coordinates {}; 

    \addplot+ [red, solid, style=thick, boxplot prepared={
        lower whisker=0.30997382763, median=0.641123, upper whisker=0.97227217236
        }] coordinates {}; 
    
    \addplot+ [red, solid, style=thick, boxplot prepared={
        lower whisker=0.22550045225785929, median=0.44583216732, upper whisker=0.6661638823874129
        }] coordinates {}; 
    
    \addplot+ [red, solid, style=thick, boxplot prepared={
        lower whisker=-0.027180170166927624, median=0.3763900441020149, upper whisker=0.7655999180371021
        }] coordinates {}; 

    \addplot+ [gray, solid, style=thick, boxplot prepared={
        lower whisker=0.3473561993167876, median=0.62772612099, upper whisker=0.9080960426800089
        }] coordinates {}; 
        
    \addplot+ [gray, solid, style=thick, boxplot prepared={
        lower whisker=0.67620528493, median=0.896537, upper whisker=1.11686871506
        }] coordinates {}; 
        
    \addplot+ [gray, solid, style=thick, boxplot prepared={
        lower whisker=-0.1797174, median=0.1812281, upper whisker=0.5421735
        }] coordinates {}; 
     
    \addplot[color=black,dashed] coordinates {
        		(0,8)
        		(0,0)
        	};
    \end{axis}
    \begin{axis}[
    scale=0.85,
    height=5.0cm,
    width=7.0cm,   
      xmax=1.2/8.5,
      xmin=-0.3/8.5,
       ymin=0,
      ymax=8,
      ytick={1,...,7},
      axis y line*=left,
      axis x line*=top,
      yticklabels={,,},
      xticklabel style={/pgf/number format/fixed},
      xlabel={Cohen's $d$}]
      \end{axis}
\end{tikzpicture}%
    \caption{}
    \label{fig5b}
\end{subfigure}%
\caption{Estimates of the ATE of stimulant medication on grade 8 tests scores, with 95\% confidence intervals: (a) numeracy test, and (b) reading test. In \textcolor{blue}{blue}: proposed TSR estimator; in \textcolor{red}{red}: alternative consistent estimators; in \textcolor{gray}{gray}: inconsistent estimators. Values are raw score points on the bottom axis and Cohen's $d$ on the top axis.}
\label{fig5}
\end{figure}
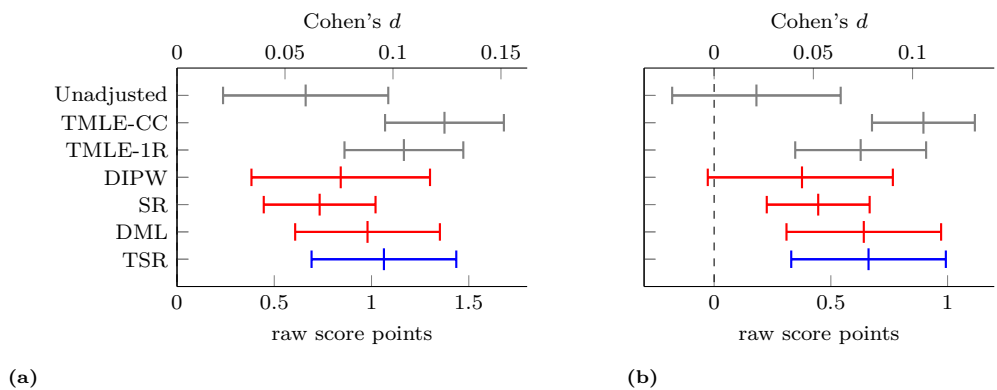

Four estimators for the ATE, each admissible with respect to the problem's $m$-graph, are computed: the DIPW, SR, and TSR estimators, as described in \cref{secPrelim,secPlugin,secidTMLE}, along with \citet{Huber} DML-based estimator. Classical sample-splitting and super-learning schemes were employed with a battery of four base algorithms: GLMs, penalized GLMs, random forest, and boosted decision trees. Results are depicted in \cref{fig5}.

All estimates agree in that they indicate a positive but small effect of stimulant medication. The TSR estimate reveals an effect of 1.06 raw score points for the numeracy test, with a 95\% confidence interval of (0.69, 1.44) and Cohen's $d$ of 0.10 (0.06, 0.13). The recovered population mean score under control is 20.21, indicating that treatment grants a 5.2\% increase in the expected test score relative to the counterfactual scenario. For the reading test, the TSR estimate yields an effect of 0.66 raw score points, with a 95\% confidence interval of (0.33, 0.99) and Cohen's $d$ of 0.08 (0.04, 0.12). With the recovered population mean score under control being 20.26, the treatment results in a 3.3\% increase in the expected test score relative to the counterfactual scenario. These findings align with accumulated clinical evidence supporting a positive but small effect of ADHD stimulant medication on academic achievement  \citep{langberg2012does, faraone2021world, Storebo2015dz,pelham2022effect,JANGMO2019423}. 

Interestingly, the TSR estimates are slightly higher than those from both the DIPW and SR approaches. This may occur when the estimate from singly-weighted IPW with mean-imputations, derived from combining $\widehat{\pi}_A$ and $\widehat{Q}_1$, produces higher estimates than both the DIPW and SR approaches, pulling the TSR estimate upward. This highlights the practical relevance of condition \textit{(iii)} in \cref{theoMR}.

The causal interpretation of these results hinges on several graphical and statistical assumptions, some of which are untestable. These include the correct specification of the $m$-graph, the implied $s$-admissibility of the chosen covariate sets, the positivity of propensity scores (also referred to as the overlap or common support assumption), and i.i.d. sampling from the target population. Furthermore, we have assumed that the employed covariates are sufficient to account for adherence and discontinuation patterns of medication over time, and that dispensation data serves as an accurate proxy for actual medication use. While the potential impacts of violations of some of these assumptions could, in principle, be explored through sensitivity analysis \citep{mathur2023,phd2} and partial identification techniques \citep{smith2019bounding}, such task is beyond the scope of this work.



\section{Conclusion and discussion}
\label{secDiscussion}

The proposed SAC serve as a method for recovering causal effects from attrition. This approach constructs and employs $s$-admissible pairs of adjustment sets that may, in principle, incorporate post-exposure and forbidden variables, yielding statistical estimands that may involve one or two sequential regression models. The SAC expands upon existing criteria that either prohibit the use of forbidden variables \citep{Correa_Tian_Bareinboim_2018,saadati2019adjustment} or permit them only under exogenous treatment assignment \citep{Mohan2014}. Compared to recent non-constructive approaches that allow the use of intermediate confounders of selection and outcome \citep{Huber}, the SAC shows advantages by identifying one or more admissible adjustment sets, as it does not restrict these sets to strict pre- and post-exposure periods. In addition, when compared to algorithmic or heuristic approaches \citep{Bhattacharya2020,tikka2021causal}, which have not been proven complete, the SAC offers a more specialized and transparent solution for the specific context of outcome attrition only. By presenting an itemized set of graphical conditions, the SAC makes assumptions about the data-generating process explicit, enabling practitioners to independently justify each component. 

We introduce the TSR estimator for cases where the recovered ATE involves two sequential regression models. The TSR estimator is built on the TMLE framework, which, given some technical conditions, demonstrates multiple robustness to model misspecification. The TSR estimator remains consistent in scenarios where both the DIPW and the naïve plug-in sequential regression approaches fail, harnessing the collective robustness conditions of regression-based, IPW-based, and imputation-based solutions. Through simulations, we compare its performance against alternative estimators. We demonstrate that, in systems where forbidden nodes directly influence selection, overlooking such post-exposure variables results in biased estimates. Furthermore, the misspecification of nuisance functions substantially affects singlyrobust estimators, even in scenarios of minimal selection bias. Finally, the TSR estimator's derived asymptotic variance estimator demonstrates relative greater robustness to model misspecification than other approaches, resulting in confidence intervals with more robust coverage probability and therefore more reliable statistical inference.

We apply these methods to real-world data to estimate the causal effect of pharmacological treatment for attention-deficit/hyperactivity (ADHD) disorder upon national test scores taken by all eligible diagnosed Norwegian schoolchildren, including those who could be exempted from taking the test. The results indicate a positive, albeit small, impact of stimulant medication on academic achievement, in line with reported clinical evidence. From a methodological standpoint, this application of the TSR procedure is justified, as clinical and contextual insights suggest that intermediate health variables may influence both academic achievement and the likelihood of exemptions or abstentions from national tests. These variables are plausibly affected by the exposure, as indicated by the reported secondary effects of common ADHD medications, such as methylphenidate-based stimulants. From a public health perspective, this inference problem is relevant, as even small rates of missing data can result in shifts in conclusions and subsequent policy designs that may not be appropriate or fully aligned with the target population.

Future work will explore appropriate criteria for comparing different minimal $s$-admissible pairs, such as by invoking the asymptotic variance of their respective plug-in estimators. Previous research in settings without attrition has shown that finding \textit{uniformly optimal} sequential adjustment sets is not always feasible \citep{Oset2,Rotnitzky}. Since our $s$-admissible pairs for attrition represent a specialized form of sequential adjustment, we hypothesize that these limitations may extend to their uniform comparison. Future research will delve deeper into these issues, exploring potential trade-offs between criteria based on optimal adjustment and on the robustness properties of the resulting semiparametric estimators.


Given the widespread challenges posed by missing data and selection bias in causal analysis, the contributions of this work provide important tools for addressing outcome attrition. We foresee extensions to settings involving multiple covariate missingness mechanisms and other forms of selection bias, and potential broader applications in areas such as data fusion and the transportability of causal relationships. 

\clearpage

\noindent\textbf{{\sffamily Acknowledgments}}: The authors are grateful to Sebastian Krumscheid and Vera Haugen Kvisgaard for their fruitful insights into the challenges of uncertainty quantification in causal inference. Special thanks are extended to Jonas Peters for his constructive feedback during the poster presentation at the Nordic Conference in Mathematical Statistics 2023 (NordStats) in Gothenburg.\\

\noindent\textbf{{\sffamily Author contributions}}: Johan de Aguas conceptualized the theoretical and methodological research questions, designed and implemented the statistical and algorithmic tools, and drafted the manuscript. Johan Pensar provided oversight on the mathematical content, particularly on the formulation of graphical criteria and theorem proving. Tomás Varnet Pérez crafted the written sections regarding the pharmacoepidemiology and public health perspectives of ADHD treatment and helped structuring the data analysis. Guido Biele secured the funding, prepossessed the data, designed the study for the application case, and provided conceptual feedback on selection bias. The final manuscript benefited from critical contributions from all authors, who have accepted responsibility for the entire content of this article.\\

\noindent\textbf{{\sffamily Funding information}}: This study was funded by the Research Council of Norway (project number 301081, PI: Guido Biele). Johan Pensar was supported by the Research Council of Norway through the centres of excellence scheme: Integreat -- The Norwegian Centre for Knowledge-driven Machine Learning (project number 332645).\\

\noindent\textbf{{\sffamily Ethical approval}}: REK ethical approval number 96604.\\

\noindent\textbf{{\sffamily Conflict of interest}}: The authors state no conflict of interest.\\

\noindent\textbf{{\sffamily Data availability statement}}: The data and code utilized for the simulation study are accessible in the personal GitHub repository at \url{https://github.com/johandh2o}. Due to the sensitive nature of the research topic, data from the application case cannot be obtained from the authors, but have to be requested from the relevant Norwegian authorities.

\clearpage


\appendix

\section{Appendix A}

\subsection{Proof of \cref{theoSAC}}
\label{appendixA1}

Let the system be described as in \cref{secProblem}, and $(W;Z)$ be an $s$-admissible pair relative to $(A;Y)$ in $\mathcal{G}$ according to  \cref{defSAC}. If $Z=\emptyset$, the graphical criteria and recovered estimand reduce to already established single-regression solutions \citep{Correa_Tian_Bareinboim_2018,Huber}. Consider now the case of $Z\neq\emptyset$, then:
\begin{align}
    & p(y\mid\doo(A=a)) & &\\
    &= \int\dd P(W)\, p(y\mid W,A=a)  &\text{ backdoor adjustment with } W&  \\
    &= \int\dd P(W)\,\int\dd Z\, p(Z,Y=y\mid W,A=a)  &\text{marginalizing out } Z&  \\
    &= \int\dd P(W)\,\int\dd Z\,\,p(Z\mid W,A=a)\, p(y\mid W,A=a,Z)  &\text{factorizing } p(Z,Y\mid W,A)&  \\
    &= \int\dd P(W)\,\int\dd P(Z\mid W,A=a)\, p(y\mid W,A=a,Z,R_Y=1)  & Y\indep R_Y\mid W,A,Z &  \\
    &= \E_W\,\E_{Z\mid W,A=a}\, p(y\mid W,A=a,Z,R_Y=1).  &\text{expectation form}&  
\end{align}
The first equality is a consequence of the first and second conditions in \cref{defSAC}. If $Y \indep_d A \mid W$ in $\mathcal{G}[A\mid Y]$, then $W$ is backdoor-admissible, even if it includes \textit{pure} descendants of $A$, because $W$ contains no forbidden nodes \citep{PearlPT}. The fourth equality follows from the third condition in \cref{defSAC}. This result implies recovery of additive effects such as the ATE. In particular, if $A$ is binary, then:
\begin{align}
    \psi &= \int  y\, \dd P(y\mid\doo(A=1)) - \int  y\, \dd P(y\mid\doo(A=0)) = \Delta_a\int y\, \dd P(y\mid\doo(A=a)) \\
    &= \Delta_a \E_W\E_{Z\mid W,A=a}\, \int  y\, \dd P(y\mid W,A=a,Z,R_Y=1)\\
    &= \E_W\Delta_a\E_{Z\mid W,A=a}\, \E\left[Y\mid W,A=a,Z,R_Y=1\right],
\end{align}

\noindent provided all expected values exist.

\subsection{Proof of \cref{lemmaCD}}
\label{appendixA2}

Let $(W;Z)$ be an $s$-admissible pair relative to $(A;Y)$ in $\mathcal{G}$, then we have that $Y\indep R_Y \mid W ,A,Z$:
\begin{align}
    \widehat{Q}_\text{CD}(W ,A) &= \int\dd \widehat{P}(Z \mid W ,A)\,\widehat{Q}_1(W ,A,Z ) & &  \\
    &= \int\dd P(Z \mid W ,A)\,\E\left[Y\mid W ,A,Z ,R_Y =1\right] &\text{correct specification of } \widehat{P}_{Z \mid W ,A}, \widehat{Q}_1&  \\
     &= \int\dd P(Z \mid W ,A)\, \E\left[Y\mid W ,A,Z \right] & Y\indep R_Y \mid W ,A,Z &  \\
     &=  \int\dd Z \, p(Z \mid W ,A)\,\int \dd Y\, Y\, p(Y\mid W ,A,Z ) &\text{integral form} & \\
     &= \int \dd Y\, Y\int\dd Z \, p(Z ,Y\mid W ,A) &\text{defactorization of } p(Z ,Y\mid W ,A) & \\
     &= \int \dd Y\, Y\, p(Y\mid W ,A) &\text{marginalizing out } Z  & \\
     &= {\E}\left[Y\mid W ,A\right]. &\text{expectation form}
\end{align}

If $\{O_i\}_{i=1}^n$ are i.i.d samples from $P_{\text{obs}}$, then, by the weak law of large numbers:
\begin{equation}
    \widehat{\psi}_{\text{CD}} = \frac{1}{n}\sum_{i=1}^{n}\Delta_a\widehat{Q}_\text{CD}(W_i,a) \overset{P_{\text{obs}}}{\longrightarrow}\int\dd P(W)\, \Delta_a{\E}\left[Y\mid W ,A=a\right] = \Delta_a{\E}\left[Y\mid \doo(A=a)\right] = \psi,
\end{equation}

\noindent where the second-to-last equality comes from ${\E}_W\E\left[Y\mid W ,A=a\right]={\E}[Y\mid \doo(A=a)]$, provided expected values exist.

\subsection{Proof of \cref{lemmaSR}}
\label{appendixA3}

Let $(W;Z)$ be an $s$-admissible pair relative to $(A;Y)$ in $\mathcal{G}$, then we have that $Y\indep R_Y \mid W ,A,Z$:
\begin{align}
    \widehat{Q}_2(W ,A) &= \widehat{\E}_{Z\mid W,A}\widehat{Q}_1(W ,A,Z ) & &  \\
     &= {\E}_{Z\mid W,A}{\E}\left[Y\mid W ,A,Z ,R_Y =1\right] &\text{correct specification of } \widehat{Q}_1,\widehat{Q}_2&  \\
     &= {\E}\left\{{\E}\left[Y\mid W ,A,Z \right]\mid W ,A\right\} & Y\indep R_Y \mid W ,A,Z &  \\
     &= {\E}\left[Y\mid W ,A\right]. &\sigma\text{-algebra } (W ,A) \subset \sigma\text{-algebra } (W ,A,Z ) & 
    \label{prove1A4}
\end{align}

If $\{O_i\}_{i=1}^n$ are i.i.d samples from $P_{\text{obs}}$, then, by the weak law of large numbers:
\begin{equation}
    \widehat{\psi}_{\text{SR}} = \frac{1}{n}\sum_{i=1}^{n}\Delta_a\widehat{Q}_\text{SR}(W_i,a) \overset{P_{\text{obs}}}{\longrightarrow}\int\dd P(W)\, \Delta_a{\E}\left[Y\mid W ,A=a\right] = \Delta_a{\E}\left[Y\mid \doo(A=a)\right] = \psi,
\end{equation}

\noindent where the second-to-last equality comes from ${\E}_W\E\left[Y\mid W ,A=a\right]={\E}[Y\mid \doo(A=a)]$, provided expected values exist.

\subsection{Proof of \cref{theoEIF}}
\label{appendixA4}

Let $(W;Z)$ be an $s$-admissible pair with $Z\neq\emptyset$ and $P$ stand for the observed data distribution. Let us express the associated $s$-formula for the ATE $\Psi[P]$ [\cref{eqSformula}] as a difference of recovered potential outcome means, so $\Psi[P] =\Delta_a \Phi^a[P]$, with:
\begin{equation}
    \Phi^a[P] = \E_{W }\E_{Z  \mid W ,A=a}\E[Y\mid W ,A=a,Z  ,R_Y=1] = \iiint \dd W \,\dd Z  \,\dd Y\, Y\, \frac{p(W )\,p(W ,a,Z  )\,p(W ,a,Z  ,1,Y)}{p(W ,a)\,p(W ,a,Z  ,1)}.
\end{equation}

Consider parametric submodel $P_\epsilon\in\mathfrak{P}$ indexed by a small fluctuation parameter $\epsilon\in\R$ and a point-mass contamination $O_i=(W_i,A_i,Z_i,R_{Y,i},Y_i^\dagger)$, such that, $P_\epsilon(O)=\epsilon\,\I({O}={O}_i)+(1-\epsilon)\,P({O})$, and with $p_\epsilon$ being its Radon-Nikodym derivative with respect to an appropriate measure $\mu$. Under some technical conditions involving \textit{(i)} fully nonparametric or saturated model $\mathfrak{P}$, \textit{(ii)} smoothness for the paths within the model, \textit{(ii)} positivity of the involved propensity scores, and \textit{(iv)} boundedness of the outcome mean, the Gâteaux derivative and their variances, we have that $\Phi^a[\cdot]$ is pathwise differentiable \citep{hines2022demystifying}. Its EIF at $P$ being evaluated at point $O_i\sim P$, denoted here $D^{a}_P(O_i)$, can be computed using the following derivation result:
\begin{equation}
    \dv{}{\epsilon}\frac{p_\epsilon(W )\, p_\epsilon(W ,a,Z  )\,p_\epsilon(W ,a,Z  ,1,Y)}{p_\epsilon(W ,a)\,p_\epsilon(W ,a,Z  ,1)}\eval_{\epsilon=0}=\sum_{j=1}^5 D_j -\frac{p(W )\, p(W ,a,Z  )\,p(W ,a,Z  ,1,Y)}{p(W ,a)\,p(W ,a,Z  ,1)},
    \label{eqAnnex1}
\end{equation}

\noindent provided such derivative exists, and where:
\begin{align}
    D_1 & = \frac{1}{p(a\mid W )\,\pr(R_Y=1\mid W ,a,Z  )}   \,\I(W =W_i,A=a,Z  =Z_i,R_Y=1,Y=Y_i), \\
    D_2 & = -\frac{p(Y\mid W ,a,Z  ,R_Y=1)}{p(a\mid W )\,\pr(R_Y=1\mid W ,a,Z  )}   \,\I(W =W_i,A=a,Z  =Z_i,R_Y=1),\\
    D_3 & = \frac{p(Y\mid W ,a,Z  ,R_Y=1)}{p(a\mid W )}   \,\I(W =W_i,A=a,Z  =Z_i), \\
    D_4 & = -\frac{p(Z  \mid W ,a)\,p(Y\mid W ,a,Z  ,R_Y=1)}{p(a\mid W )}   \,\I(W =W_i,A=a), \\
    D_5 & = p(Z  \mid W ,a)\,p(Y\mid W ,a,Z  ,R_Y=1)  \,\I(W =W_i).
\end{align}

By interchanging the order of integration and differentiation:
\begin{align}
      D^{a}_P(O_i) &:= \dv{}{\epsilon}\Phi^a[P_\epsilon]\eval_{\epsilon=0} = \iiint \dd W \,\dd Z  \,\dd Y\, Y\, \sum_{j=1}^5 D_j -\Phi^a [P]\\ 
    &= \frac{\I(A_i=a,R_{Y,i}=1)\, (Y_i-{\E}[Y\mid W_i,A_i,Z_i,R_Y=1])}{p(A_i\mid W_i)\,\pr(R_Y=1\mid W_i,A_i,Z  )}\\
    &\, + \frac{1}{p(A_i\mid W_i)}\left({\E}[Y\mid W_i,A=a,Z_i,R_Y=1]-\E_{Z  \mid W_i,A=a} {\E}[Y\mid W_i,A=a,Z  ,R_Y=1] \right)\\
    &\,+ \E_{Z  \mid W_i,A=a} {\E}[Y\mid W_i,A=a,Z  ,R_Y=1] -\Phi^a [P].
\end{align}

\noindent provided positivity conditions: $0<\pr(A=1\mid W)< 1$ and $0<\pr(R_Y=1\mid W,A,Z)< 1$, the existence of expected values, and bounded variance of $D^a_P(O)$. This result leads directly to the EIF of the $s$-formula for the ATE $\Psi[\cdot]$ at $P$ evaluated at point $O_i\sim P$:
\begin{align}
    D^\Psi_P(O_i) &= D^1_P(O_i) - D^0_P(O_i) = D^Y_P(O_i) +D^Z_P(O_i)+ D^W_P(O_i)\\
    &=\frac{A_i-\pi_A(W_i)}{\pi_A(W_i)[1-\pi_A(W_i)]}\frac{R_{Y,i}}{\pi_R(W_i,A_i,Z_i)}\,[Y_i-{Q}_1(W_i,A_i,Z_i)]\\
    &\quad +\frac{A_i-\pi_A(W_i)}{\pi_A(W_i)[1-\pi_A(W_i)]}\left[{Q}_1(W_i,A_i,Z_i)- {Q}_2(W_i,A_i) \right]\\
    &\quad +\Delta_a {Q}_2(W_i,a) -\psi,
\end{align}

\noindent where $\pi_A(W) \!=\! \pr(A=1\mid W)$, $\pi_R(W,A,Z) \!=\! \pr(R_Y=1\mid W,A,Z)$ are, respectively, the exposure and selection propensity scores, ${Q}_1(W,A,Z) \!=\! {\E}\left[Y\mid W,A,Z,R_Y=1\right]$, ${Q}_2(W,A) \!=\!{\E}_{Z\mid W,A}{Q}_1(W,A,Z)$, and $\psi \!=\! \Psi[P]$.

\subsection{Proof of \cref{theoMR}}
\label{appendixA5}

Let $\widetilde{D}$ be the function given in \cref{eqEIF1,eqEIF2,eqEIF3,eqEIF4} but replacing $({Q}_1,{Q}_2,{\pi}_A,{\pi}_R)$ for the putative models $(\widetilde{Q}_1,\widetilde{Q}_2,\widetilde{\pi}_A,\widetilde{\pi}_R)$. Then:
\begin{align} \label{MR1}
\E \widetilde{D}(O) &= \E_ W \frac{\pi_A(W)}{\widetilde{\pi}_A(W)}\E_{ Z\mid W,A=1}\frac{\pi_R( W,1, Z)}{\widetilde{\pi}_R( W,1, Z)}\left[{Q}_1( W,1, Z)-\widetilde{Q}_1( W,1, Z)\right] \\  \label{MR2}
&- \E_ W\frac{1-\pi_A(W)}{1-\widetilde{\pi}_A(W)}\E_{ Z\mid W,A=0}\frac{\pi_R( W,0, Z)}{\widetilde{\pi}_R( W,0, Z)}\left[{Q}_1( W,0, Z)-\widetilde{Q}_1( W,0, Z)\right]  \\  \label{MR3}
&+ \E_ W \frac{\pi_A(W)}{\widetilde{\pi}_A(W)} \E_{ Z\mid W,A=1}\widetilde{Q}_1( W,1, Z) - \E_ W \frac{1-\pi_A(W)}{1-\widetilde{\pi}_A(W)} \E_{ Z\mid W,A=0}\widetilde{Q}_1( W,0, Z)  \\ \label{MR4}
&-\left[ \E_ W \frac{\pi_A(W)}{\widetilde{\pi}_A(W)} \widetilde{Q}_2( W,1) - \E_ W \frac{1-\pi_A(W)}{1-\widetilde{\pi}_A(W)} \widetilde{Q}_2( W,0)\right]   \\  \label{MR5}
&+ \E_W\Delta_a \widetilde{Q}_2( W,a)-\psi.
\end{align}

Trivially, $\E \widetilde{D}(O)=0$ when $(\widetilde{Q}_1,\widetilde{Q}_2,\widetilde{\pi}_A,\widetilde{\pi}_R)=({Q}_1,{Q}_2,{\pi}_A,{\pi}_R)$. To prove that the EIF defines a robust estimating equation, we have to show that it is zero even when some putative models are misspecified.

\textbf{Case 1: $\widetilde{Q}_1,\widetilde{Q}_2$ are correctly specified}: If $\widetilde{Q}_1=Q_1$ and $\widetilde{Q}_2=Q_2$, terms in \eqref{MR1},  \eqref{MR2} and  \eqref{MR5} are exactly zero. Moreover, $\E_{ Z\mid W,A=a}\widetilde{Q}_1( W,a, Z)= \E_{ Z\mid W,A=a}{Q}_1( W,a, Z)  = {Q}_2( W,a)=\widetilde{Q}_2( W,a)$, making the sum of the remaining terms null.

\textbf{Case 2: $\widetilde{\pi}_A,\widetilde{\pi}_R$ are correctly specified}: If $\widetilde{\pi}_A={\pi}_A$ and $\widetilde{\pi}_R={\pi}_R$, all ratios ${\pi}_A(W)/\widetilde{\pi}_A(W)$,  $[1-{\pi}_A(W)]/[1-\widetilde{\pi}_A(W)]$,  ${\pi}_R(W,A,Z)/\widetilde{\pi}_R(W,A,Z)$ are equal to one. The sum of the first components in terms \eqref{MR1} and \eqref{MR2} evaluates to $\psi$, canceling out the second component in term  \eqref{MR5}, while the sum of their respective second components cancels out term  \eqref{MR3}. Term  \eqref{MR4} evaluates to $-\E_W\Delta_a \widetilde{Q}_2( W,a)$, rendering its addition to the first component of term  \eqref{MR5} null.

\textbf{Case 3: $\widetilde{Q}_1,\widetilde{\pi}_A$ are correctly specified}: If $\widetilde{Q}_1=Q_1$ and $\widetilde{\pi}_A={\pi}_A$, terms  \eqref{MR1} and  \eqref{MR2} are precisely zero, and ratios ${\pi}_A(W)/\widetilde{\pi}_A(W)$,  $[1-{\pi}_A(W)]/[1-\widetilde{\pi}_A(W)]$ are equal to one. Term  \eqref{MR3} evaluates to $\psi$, and term  \eqref{MR4} evaluates to $-\E_W\Delta_a \widetilde{Q}_2( W,a)$. Both effectively cancel out term  \eqref{MR5}.

For clarity and to avoid cluttered notation, sequential expectations are here presented in a streamlined notation without nested brackets. Therefore, in all the aforementioned expressions, terms such as\\ $\E_{X_1} f_1(X_1)\E_{X_2\mid X_1} f_2(X_1,X_2)$ must be interpreted as $\E\left\{ f_1(X_1)\cdot\E\left[ f_2(X_1,X_2)\mid X_1\right]\right\}$.

\subsection{Proof of linearity remark}
\label{appendixA6}

Let $P$ be a convex combination of two compatible observed data distributions $P',P''\in\mathfrak{P}$, so there is $\alpha\in[0,1]$ such that $P=\alpha P' +(1-\alpha)P''$. Let $P',P''\in\mathfrak{P}$ agree on the marginals $P_W$ and $P_{Z\mid W,A}$ almost surely. Then:
\begin{align}
    \Psi[P] &= \E_W\Delta_a\E_{Z\mid W, A=a}\int  y\, \dd P(y\mid W,A=a,Z,R_Y=1)\\
    &= \E_W\Delta_a\E_{Z\mid W, A=a}\int  y\, \dd (\alpha P'(y\mid W,A=a,Z,R_Y=1)+(1-\alpha)P''(y\mid W,A=a,Z,R_Y=1))\\
     &= \alpha\E_W\Delta_a\E_{Z\mid W, A=a}\int  y\, \dd  P'(y\mid W,A=a,Z,R_Y=1)\\
     &\quad +(1-\alpha)\E_W\Delta_a\E_{Z\mid W, A=a}\int  y\, \dd  P''(y\mid W,A=a,Z,R_Y=1)\\
     &=\alpha \E_W\Delta_a\E_{Z\mid W, A=a}Q_1'(W,a,Z) + (1-\alpha) \E_W\Delta_a\E_{Z\mid W, A=a}Q_1''(W,a,Z)\\
     &=\alpha \Psi[P'] + (1-\alpha)\Psi[P'']
\end{align}

Thus, $\Psi$ is linear in $P(Y\mid W,A,Z,R_Y=1)$. We say it is linear in $Q_1$. Now let $P',P''\in\mathfrak{P}$ agree on the marginals $P_W$ and $P_{Y\mid W,A,Z}$ instead, almost surely. Then:
\begin{align}
    \Psi[P] &= \E_W\Delta_a\int \dd P(Z\mid W, A=a) Q_1(W,a,Z)\\
    &= \E_W\Delta_a\int \dd ( \alpha P'(Z\mid W, A=a)  + (1-\alpha)P''(Z\mid W, A=a) ) Q_1(W,a,Z)\\
     &= \alpha\E_W\Delta_a\int \dd P'(Z\mid W, A=a) Q_1(W,a,Z)\\
     &\quad +(1-\alpha)\E_W\Delta_a\int \dd P''(Z\mid W, A=a) Q_1(W,a,Z)\\
     &=\alpha \E_W\Delta_a Q_2'(W,a) + (1-\alpha) \E_W\Delta_a Q_2''(W,a)\\
     &=\alpha \Psi[P'] + (1-\alpha)\Psi[P'']
\end{align}

Thus, $\Psi$ is linear in $P(Z\mid W,A)$ given $P(Y\mid W,A,Z,R_Y=1)$. We say it is linear in $Q_2$ given $Q_1$.

\subsection{Conditions for vanishing second-order remainder}
\label{appendixA7}

Let $\Omega$ be the second-order remainder from using updated estimator $\psi^*=\Psi[P^*]$ for $\psi$, then:
\begin{align}
    \Omega &= \psi^*-\psi+\E D_{P^*}^\Psi(O) = \sum_{j=1}^5\Omega_j\\
    \Omega_1 &:= \E_{W,Z,R_Y,Y\mid A=1}\left[\frac{\pi_A(W)\, R_Y}{\widehat{\pi}_A(W)\,\widehat{\pi}_R(W,1,Z)}Y \right] - \E_{W,Z,R_Y,Y\mid A=0}\left[\frac{[1-\pi_A(W)]\, R_Y}{[1-\widehat{\pi}_A(W)]\,\widehat{\pi}_R(W,0,Z)}Y\right],\\
     \Omega_2 &:= -\E_{W,Z\mid A=1}\left[\frac{\pi_A(W)\,\pi_R(W,1,W)}{\widehat{\pi}_A(W)\,\widehat{\pi}_R(W,1,Z)}Q_1^*(W,1,Z) \right]\\  
     &\quad + \E_{W,Z\mid A=0}\left[\frac{[1-\pi_A(W)]\,\pi_R(W,0,W)}{[1-\widehat{\pi}_A(W)]\,\widehat{\pi}_R(W,0,Z)}Q_1^*(W,0,Z)\right],\\
    \Omega_3 &:= \E_{W,Z\mid A=1}\left[\frac{\pi_A(W)}{\widehat{\pi}_A(W)}Q_1^*(W,1,Z) \right]-\E_{W,Z\mid A=0}\left[\frac{1-\pi_A(W)}{1-\widehat{\pi}_A(W)}Q_1^*(W,0,Z)\right],\\
    \Omega_4 &:= -\E_W\left[\frac{\pi_A(W)}{\widehat{\pi}_A(W)}Q_2^*(W,1) -\frac{1-\pi_A(W)}{1-\widehat{\pi}_A(W)}Q_2^*(W,0)\right],\\
    \Omega_5 &:= \E_W\left[\Delta_a Q_2^*(W,a) - \Delta_a Q_2(W,a)\right].
\end{align}

Given $s$-admissible $(W;Z)$, we have the following equivalences for treatment arm $A=1$ given $R_Y=1$:
\begin{align}
    \E_{W,Z,R_Y,Y\mid A=1}\left[\frac{\pi_A(W)\, R_Y}{\widehat{\pi}_A(W)\,\widehat{\pi}_R(W,1,Z)}Y \right] &=\E_{W,Z\mid A=1}\left[\frac{\pi_A(W)}{\widehat{\pi}_A(W)\,\widehat{\pi}_R(W,1,Z)}\,\E_{Y,R\mid W,A=1,Z}[R_Y\,Y]\right],\\
    &=\E_{W,Z\mid A=1}\left[\frac{\pi_A(W)\,{\pi}_R(W,1,Z)}{\widehat{\pi}_A(W)\,\widehat{\pi}_R(W,1,Z)}Q_1(W,1,Z) \right],\\
        \E_{W,Z\mid A=1}\left[\frac{\pi_A(W)}{\widehat{\pi}_A(W)}Q_1^*(W,1,Z) \right] &=\E_W\left[\frac{\pi_A(W)}{\widehat{\pi}_A(W)}\E_{Z\mid W,A=1}\widehat{Q}^*_1(W,1,Z) \right],\\
        &=\E_W\left[\frac{\pi_A(W)}{\widehat{\pi}_A(W)}Q_2(W,1)\right].
\end{align}

The latter can be justified by invoking the definition of $Q_2$, or by considering $\E_{Z\mid W,A}\widehat{Q}^*_1(W,A,Z)$ to be consistent for $Q_2(W,A)$. Thus, by leveraging similar equivalences for treatment arm $A=0$ given $R_Y=1$:
\begin{align}
\sum_{j=1}^2\Omega_j &=\E_{W,Z\mid A=0}\left[\frac{1-{\pi}_A(W)}{1-\widehat{\pi}_A(W)}\,\frac{\pi_R(W,0,Z)}{\widehat{\pi}_R(W,0,Z)}[Q_1^*(W,0,Z)-Q_1(W,0,Z)] \right]\\
&\,+ \E_{W,Z\mid A=1}\left[\frac{{\pi}_A(W)}{\widehat{\pi}_A(W)}\,\frac{\pi_R(W,1,Z)}{\widehat{\pi}_R(W,1,Z)}[Q_1(W,1,Z)-Q_1^*(W,1,Z)] \right],\\
    \sum_{j=3}^5\Omega_j &= \E_W\left[\frac{\widehat{\pi}_A(W)-\pi_A(W)}{1-\widehat{\pi}_A(W)}[Q_2^*(W,0)-Q_2(W,0)] \right]\\
    &\,+ \E_W\left[\frac{\widehat{\pi}_A(W)-\pi_A(W)}{\widehat{\pi}_A(W)}[Q_2^*(W,1)-Q_2(W,1)] \right].
\end{align}

Then, by Cauchy-Schwarz's and Minkowski's inequalities of the $L^2(P)$ norm, and under positivity and consistency of  nuisance parameters and their estimators:
\begin{align}
\sum_{j=1}^2\Omega_j &\leq \Biggl(\underbrace{\norm{\frac{{\pi}_A(W)}{\widehat{\pi}_A(W)}-1}_2}_{o_P(n^{-1/4})}+1\Biggr)\Biggl(\underbrace{\norm{\frac{{\pi}_R(W,A,Z)}{\widehat{\pi}_R(W,A,Z)}-1}_2}_{o_P(n^{-1/8})}+1\Biggr)\underbrace{\norm{Q_1^*(W,A,Z)-Q_1(W,A,Z)}_2}_{o_P(n^{-1/8})}\\
\sum_{j=3}^5\Omega_j &\leq \underbrace{\norm{\frac{{\pi}_A(W)}{\widehat{\pi}_A(W)}-1}_2}_{o_P(n^{-1/4})}\underbrace{\norm{Q_2^*(W,A)-Q_2(W,A)}_2}_{o_P(n^{-1/4})}. 
\end{align}

Hence, $\Omega\in o_P(n^{-1/2})$.

\subsection{Simulation setup I}
\label{appendixA8}

We generate 5\,000 i.i.d. samples from the following SCM, with associated graph presented in \cref{fig3a}:
\begin{align}
W_1 &\sim N(0,1), &\\
A &= \mathbb{I}[0.90\,W_1 -0.09\,\text{sign}(W_1)\,W_1^2 + U_A > 0],\quad & U_A\sim N(0,1),\\
Z_1 &= -0.50 + A + U_{Z_1},\quad & U_{Z_1}\sim N(0,1),\\
Z_2 &=  0.20\,[4.0 + 0.05\,(2A-1) + 0.50\,Z_1 + 0.05\,(2A-1)\, Z_1 + U_{Z_2}]^2,\quad & U_{Z_2}\sim N(0,1),\\
Y &=  3.00\,W_1+ 1.50\sqrt{\abs{W_1}} - 0.25\,(2A-1) + 0.50\,(2A-1)\, W_1   & \\
 &\qquad + 1.25\,Z_1 + 0.25\,(2A-1)\, Z_1 + Z_2 + 0.50\,(2A-1)\, Z_2+ U_Y,\quad & U_Y\sim N(0,7),
\end{align}

\noindent where $N(0,\sigma)$ is the Gaussian distribution with zero mean and standard deviation $\sigma$. The selection mechanism has a probit specification depending on mediators $Z_1$ and $Z_2$.
\begin{equation}
R = \mathbb{I}[\theta + 0.29\,Z_1 + 0.54\,Z_2  + U_R > 0],\quad U_R\sim N(0,1),
\end{equation}

\noindent where parameter $\theta$ controls the base missingness rate: $\theta=-1.90$ for 50\% rate, $\theta=-0.90$ for 25\% rate, and $\theta=-0.30$ for 15\% rate. Oracle specification produces McFadden's pseudo-$R^2$ in the range 17\%---38\% across various models and scenarios. This reflects data with a moderate level of noise, as frequently observed in applied research. 

Misspecification is introduced by substituting the super-learning schemes of nuisance parameters with the following parametric specifications: \textit{(i)} a linear probability model for $\widehat{\pi}_R$ given $Z_1^2$; \textit{(ii)} a linear regression model for $\widehat{Q}_1$ given $(W^2,A,AW,Z_1)$; \textit{(iii)} a linear regression model for $\widehat{Q}_2$ given $((W-1.7)^2,A)$; and \textit{(iv)} a linear probability model for $\widehat{\pi}_A$ given $U_0\sim N(0,1)$ unrelated to the other variables in the SCM. For the approaches involving a single-regression  (TMLE-1R and TMLE-CC), the former two are instead: \textit{(i)} a linear probability model for $\widehat{\pi}_R$ given $W_1^2$ and \textit{(ii)} a linear regression model for $\widehat{Q}_1$ given $(W^2,A,AW)$.

\subsection{Simulation setup II}
\label{appendixA9}

We generate 5\,000 i.i.d. samples from the following SCM, with associated graph presented in \cref{fig2a}:
\begin{align}
U_1 &\sim N(0,1),\\
B_1 &\sim N(0,1),\\
A &= \mathbb{I}[0.90\,B_1 -0.10\,\text{sign}(B_1)\,B_1^2 + U_A > 0],\quad & U_A\sim N(0,1),\\
Y &=  2.00\,B_1+ \text{sign}(B_1)\,B_1^2 -1.20\,(2A-1)+ 0.50\,(2A-1)B_1   & \\
 &\qquad + 2.00\,U_1+ 1.20\,(2A-1)U_1,\quad & U_Y\sim N(0,7),\\
C_1 &= 0.20\,B_1^2+ 0.50\,(2A-1)+ 0.20\,(2A-1)B_1,\quad & U_{C_1}\sim N(0,1),\\
C_2 &=  0.10\,U_1+ 0.10\,C_1 +0.10\,U_1C_1\quad & U_{C_2}\sim N(0,2.5),\\
R &= \mathbb{I}[0.2 + 0.2\,(2A-1)+ 0.1\,C_2 -0.1\,(2A-1)C_2  + U_R > 0]\quad & U_R\sim N(0,0.7).
\end{align}

Oracle specification produces McFadden's pseudo-$R^2$ in the range 20\%---30\% across various models. Misspecification is introduced by substituting the super-learning schemes of $\widehat{Q}_2$ for a linear regression model  given $(\abs{W-1},A)$.

\clearpage

\vskip 0.2in
\bibliographystyle{myunsrtnat}
\bibliography{references}

\end{document}